\documentclass[letterpaper,11pt]{article}
\pdfoutput=1
\usepackage{hlmacro} 
\usepackage{jheppub} 

\usepackage{tabularray}
\usepackage{booktabs}
\usepackage[english]{babel}
\usepackage{amsmath,amssymb,amstext, amsfonts}
\usepackage{hyperref}
\usepackage{graphicx}
\usepackage{subcaption}
\usepackage{amsthm}
\usepackage{bbm}
\usepackage{upgreek}
\usepackage{framed} 
\usepackage{multirow}
\usepackage[utf8x]{inputenc}
\usepackage{selinput}
\usepackage{bm}
\usepackage{float}
\usepackage{yfonts}
\usepackage{mathtools}
\usepackage{enumitem}   
\usepackage{longtable}
\usepackage{tabularx}
\usepackage{slashed}
\usepackage{yhmath}
\usepackage{afterpage}
\usepackage{fontawesome}
\usepackage{colortbl}
\usepackage{tcolorbox}

\usepackage{tikz}
\usetikzlibrary{arrows,chains,matrix,positioning,scopes}
\usepackage{tikz-cd}
\usetikzlibrary{matrix, calc, arrows}

\makeatletter
\newcommand{\github}[1]{%
   \href{#1}{\faGithub}%
}
\makeatother

\makeatletter
\g@addto@macro\bfseries{\boldmath}
\makeatother

\def \O {\cO}

\def\wt{\widetilde}
\def\wh{\widehat}

\def \cA {{\cal A}}

\def \cC {{\cal C}}
\def \cD {{\cal D}}
\def \cE {{\cal E}}
\def \cF {{\cal F}}

\def \cL {{\cal L}}
\def \O {{\cal O}}
\def \cO {{\cal O}}
\def \cP {{\cal P}}

\def \cX {{\cal X}}

\def \tDelta {{\tilde \Delta}}

\newcommand{\eec}[2]{\eps_{#1}\cdot\eps_{#2}} 
\newcommand{\EEc}[2]{\cE_{#1}\cdot\cE_{#2}}
\newcommand{\ddc}[2]{\eps_{#1}\cdot p_{#2}} 
\newcommand{\DDc}[2]{\cE_{#1}\cdot\cX_{#2}}
\newcommand{\DDtc}[2]{\cE_{#1}\cdot\cP_{#2}}

\newcommand{\ee}[1]{\eps_{#1}} 
\newcommand{\pp}[1]{p_{#1}} 
\newcommand{\EE}[1]{\cE_{#1}}
\newcommand{\dd}[1]{d_{#1}} 
\newcommand{\DD}[1]{\cD_{#1}} 
\newcommand{\FF}[1]{\cF_{#1}} 
\newcommand{\DDt}[1]{\wt\cD_{#1}} 
\newcommand{\PP}[1]{\cP_{#1}} 
\renewcommand{\XX}[1]{\cX_{#1}} 

\title{Amplitude Basis for Conformal Correlators}

\author{Hayden Lee$^1$ and Xinkang Wang$^2$} 

\affiliation{$^1$ Kavli Institute for Cosmological Physics, University of Chicago, Chicago, IL 60637, USA}
\affiliation{$^2$ Department of Physics, University of Chicago, Chicago, IL 60637, USA}
\emailAdd{haydenl@uchicago.edu, xinkangwang@uchicago.edu} 
 
\abstract{We present a classification of conformally-invariant three-point tensor structures in $d$ dimensions that parallels the classification of three-particle scattering amplitudes in $d+1$ dimensions. 
Using a set of canonically-normalized weight-shifting operators, we construct a basis of three-point structures involving conserved currents or stress tensors and non-conserved spinning operators, directly from their amplitude counterparts. 
As an application, we also examine the conformal block expansion of the four-point functions of external currents and stress tensors in this amplitude basis.
Our results can be useful for conformal bootstrap applications involving spinning correlators as well as Witten diagram computations in anti-de Sitter space.
}

\preprint{}

\makeatletter
\def\@fpheader{\ }
\makeatother

\begin{document}
\maketitle
\flushbottom
\newpage

\section{Introduction}
\setlength\parskip{4pt}

A conformal field theory (CFT) is characterized non-perturbatively by a set of numbers specifying the spectrum of local operators---scaling dimensions $\Delta_i$ and spins $\ell_i$---along with the coefficients of their three-point functions.
Thanks to the convergence properties of operator product expansion (OPE), these CFT data determine not only two- and three-point functions, but are sufficient to determine all higher-point functions through an expansion in conformal blocks.
The modern conformal bootstrap program~\cite{Rattazzi:2008pe}, building upon the foundational work of~\cite{Polyakov:1974gs,Belavin:1984vu}, reformulates OPE consistency in two different channels as an infinite set of constraints that can be systematically studied numerically.
The past 15 years have seen an explosion of progress on the bootstrap approach to CFTs on both numerical and analytic fronts, see~\cite{Poland:2018epd,Poland:2022qrs,Rychkov:2023wsd,Bissi:2022mrs,Hartman:2022zik} for reviews.  

So far the predominant focus of bootstrap studies has centered on scalar four-point functions, and even from this relatively narrow set of constraints surprisingly strong results have emerged. 
There are, however, good reasons to expand these investigations to include external operators with nonzero spin, such as global symmetry currents and stress tensors.
Notably, stress tensors are ubiquitous in any local CFTs and couple to all operators that are singlets under global symmetries.
Given their distinctive characteristics, spinning correlators can provide a unique lens into CFT dynamics and help improve the bootstrap constraints obtained solely from scalar correlators.
Recent advancements in this direction include bootstrapping three-dimensional CFT data using the four-point functions of global symmetry currents \cite{Dymarsky:2017xzb, Reehorst:2019pzi, He:2023ewx} and stress tensors~\cite{Dymarsky:2017yzx}.

Dealing with spinning correlators presents an extra layer of technicality compared to scalar cases due to the challenges of managing various tensor structures.
The first task in this business is to classify independent tensor structures for conformal three-point functions, which serve as building blocks for higher-point functions.
The three-point function of spinning operators $\O_i$ takes the general form
\begin{align}
	\LA \O_1\O_2\O_3\RA = \sum_a \lambda_{\O_1\O_2\O_3}^{(a)}\LA \O_1\O_2\O_3\RA^{(a)}\,,\label{threeptexp}
\end{align}
where $\lambda_{\O_1\O_2\O_3}^{(a)}$ are the coefficients of some conformally-invariant tensor structures $\LA \O_1\O_2\O_3\RA^{(a)}$ labeled by $a$.
We are especially interested in cases where some of these operators are conserved.
The immediate question that arises is: what basis of independent tensor structures should we work with? 
The most basic approach is to write down a general ansatz with an over-complete set of some conformally-invariant building blocks, and then determine relations among their coefficients by imposing conservation, leaving a few independent tensor structures in the end.
However, a naive such procedure can lead to unnecessarily complex algebra and dependencies on OPE data, complicating practical bootstrap applications.  
To address these complexities, recent years have seen significant progress in developing novel approaches for dealing with spinning conformal correlators, see~\cite{Costa:2011dw,Costa:2011mg,Giombi:2011rz,Zhiboedov:2012bm, Dymarsky:2013wla,Costa:2014kfa,Costa:2014rya,Penedones:2015aga, Iliesiu:2015akf,Rejon-Barrera:2015bpa,CastedoEcheverri:2015mkz,Costa:2016xah,Costa:2016hju,CastedoEcheverri:2016dfa, Kravchuk:2016qvl,Schomerus:2016epl,Sleight:2017fpc,Schomerus:2017eny,Kravchuk:2017dzd,Cuomo:2017wme,Karateev:2018oml,Fortin:2020des,Caron-Huot:2021kjy}.

A few notable classifications of conformally-invariant three-point structures were introduced in~\cite{Kravchuk:2016qvl,Karateev:2018oml,Caron-Huot:2021kjy}. 
Their main idea involved ``gauge-fixing'' the conformal symmetry by going to the conformal frame $(x_1,x_2,x_3)=(0,x,\infty)$, and then organizing the tensor structures through the ``little group,'' consisting of transformations that leave this configuration invariant.
This group-theoretical approach highlights the fact that the counting of correlators in CFT$_d$ matches that of on-shell scattering amplitudes in QFT$_{d+1}$~\cite{Costa:2011mg,Kravchuk:2016qvl}.
In particular, by exploiting the property of massless particles having two helicity states in QFT$_4$, the basis for CFT$_3$ constructed in~\cite{Caron-Huot:2021kjy} fully diagonalizes the OPE data for mean-field correlators of conserved currents.

These constructions motivate the idea of organizing tensor structures in CFT$_d$ using amplitude structures in QFT$_{d+1}$ as a convenient kinematical basis, where the latter can effectively be analyzed through correlators in $(d+1)$-dimensional anti-de Sitter space (AdS).
Indeed, the study of spinning conformal correlators fits naturally in the context of the AdS/CFT correspondence~\cite{Maldacena:1997re, Witten:1998qj, Gubser:1998bc}, with conserved currents $J$ and stress tensors $T$ on the boundary being dual to photons (or gluons) and gravitons in the bulk AdS, respectively. 
Moreover, non-conserved spinning operators, which generically appear in the OPE of conserved tensors, are dual to massive spinning particles in AdS.
As such, it is useful to perform a systematic construction of AdS three-point functions involving massless and massive spinning particles, which reduce to three-particle amplitudes in the flat-space limit.

A key advancement in computing spinning conformal correlators has been the development of 
the embedding-space formalism~\cite{Costa:2011dw,Costa:2014kfa} and weight-shifting operators~\cite{Karateev:2017jgd}. 
These tools have enabled an efficient computation of spinning correlators by applying differential operators to simpler scalar seeds, and had important applications in CFT~\cite{Karateev:2018oml,Kravchuk:2018htv,Costa:2018mcg} and in cosmology~\cite{Arkani-Hamed:2018kmz, Baumann:2019oyu, Baumann:2020dch}. 
Recently, the bulk interpretation of weight-shifting operators was elucidated in~\cite{Lee:2022fgr, Li:2022tby}, where they were identified as the AdS analogs of on-shell kinematic building blocks for flat-space amplitudes, given appropriate normalization and ordering.
With this new understanding, it becomes possible to directly construct AdS correlators by substituting these kinematic building blocks from a given amplitude. 
In particular, this suggests that massless-massless-massive amplitudes in flat space, which have been fully classified in~\cite{Chakraborty:2020rxf}, can be directly ``uplifted'' to AdS.

In this paper, we provide a systematic approach to constructing three-point tensor structures in CFT$_d$ by exploiting the known classification of three-particle amplitudes in QFT$_{d+1}$. 
Specifically, we focus on parity-even, three-point functions of one or two conserved operators with non-conserved, totally symmetric spin-$\ell$ operators~$\O_\ell$, such as $\LA JJ\O_\ell \RA$, $\LA TT\O_\ell \RA$, $\LA J\O_\ell\O_\ell \RA$, and $\LA T\O_\ell\O_\ell \RA$.\footnote{These three-point functions involving higher-spin operators play an important role in analytic studies of conformal collider and causality bounds~\cite{Hofman:2008ar,Camanho:2014apa,Li:2015itl,Afkhami-Jeddi:2016ntf, Cordova:2017zej, Costa:2017twz,Meltzer:2017rtf,Afkhami-Jeddi:2018apj,Meltzer:2018tnm,Gillioz:2018kwh,Kaplan:2019soo}.}
As an application, we also derive the corresponding differential representations for the conformal blocks of external conserved currents and stress tensors due to totally symmetric tensor exchanges in $d\ge 3$. 

The paper is organized as follows. In Section~\ref{sec:gen}, we introduce the differential building blocks for conformal correlators, including scalar seed functions and weight-shifting operators. We also describe a procedure of uplifting scattering amplitudes to AdS using these basic ingredients.
In Section~\ref{sec:ampbasis}, we present a basis of tensor structures for conserved-conserved-(non-conserved) and conserved-(non-conserved)-(non-conserved) correlators. These are constructed using the same classification as massless-massless-massive and massless-massive-massive scattering amplitudes.
In Section~\ref{sec:spinningblocks}, we present the corresponding differential representations for spinning conformal blocks with external conserved currents and stress tensors. 
We conclude in Section~\ref{sec:con}. 
In Appendix~\ref{app:bulk}, we consider the construction of differential representations from the bulk perspective. 
In Appendix~\ref{app:pairing}, we compare our results with the helicity basis constructed in~\cite{Caron-Huot:2021kjy}.

\section{Differential Building Blocks}\label{sec:gen}

We begin with an overview of the differential operator technology in conformal field theory.
In~\S\ref{sec:seed}, we introduce scalar seed three-point functions, followed by the presentation of the relevant weight-shifting operators in~\S\ref{sec:WS}. 
We then describe a prescription for uplifting flat-space amplitudes to AdS in \S\ref{sec:amptocorr}.

\subsection{Three-Point Scalar Seeds}\label{sec:seed}

A natural language to describe conformal correlators is the embedding-space formalism~\cite{Dirac:1936fq,Costa:2011mg}, where conformal transformations on $\mathbb{R}^d$ are realized as Lorentz transformations on a projective lightcone embedded in $\mathbb{R}^{1,d+1}$. 
We will denote coordinates in embedding space by~$X^A$, whose indices $A=0,1,\cdots,d+1$ can be raised and lowered by the Minkowski metric $\eta_{AB}$. Because embedding-space coordinates have $d+2$ components, we impose two constraints on the coordinates in order to get the correct number of degrees of freedom for position-space coordinates $x^i$ in $\mathbb{R}^d$, with $i=1,\cdots,d$. 
This can be achieved by restricting to the Euclidean section of the projective lightcone, defined by the constraints $X^2 =0$ and $X^+=1$, where $X^\pm = X^0 \pm X^{d+1}$ are the lightcone coordinates. The Euclidean section is parameterized by
\begin{align}
	X^A = (X^+, X^-, X^i ) = (1,\, x^2,\, x^i)\,.
\end{align}
When analyzing tensors in embedding space, it is useful to adopt an index-free notation and contract all tensor indices with auxiliary null polarization vectors $Z^A$, with $Z^2=0$ and $X\cdot Z=0$. It is parameterized on the Euclidean section by
\begin{align}
	Z^A = (Z^+,Z^-,Z^i ) = (0,\,  x\cdot z,\,  z^i)\,,
\end{align}
where $z^i$ is the corresponding auxiliary null vector in $\mathbb{R}^d$, with $z^2 =0$.

Since a scalar three-point function of given scaling dimensions is uniquely fixed by conformal symmetry up to normalization, we can equivalently compute this correlator from the bulk AdS.
The bulk-to-boundary propagator of a scalar field in embedding space is
\begin{align}
	\Pi_\Delta(X,Y) = \frac{N_{\Delta}}{(-2X\cdot Y)^\Delta}\,,\quad N_\Delta \equiv \frac{\Gamma(\Delta)}{(2\pi)^{\frac{d+1}{2}}}\,,\label{Piscalar}
\end{align}
where $Y$ denotes the bulk coordinate in AdS$_{d+1}$, with $Y^0>0$ and $Y^2=-1$ in units where the AdS radius is set to one.
Note that this normalization is different from the standard convention, but it is the one that most directly relates to flat-space amplitude structures in momentum space.\footnote{
Given that scattering amplitudes live in momentum space, this comparison is naturally done for momentum-space correlators. In momentum space, the normalization choice \eqref{Piscalar} ensures a unit normalization of an $n$-point scalar seed function in the flat-space limit~\cite{Lee:2022fgr}.
If desired, switching to the standard normalization in embedding space is straightforward, which might be useful if one is interested in computing AdS correlators with precise coupling constants (see 
\cite{Li:2022tby} and Appendix~\ref{app:bulk}). 
This will change both the normalization of the scalar seeds and that of the weight-shifting operators that we discuss next.
Our primary interest in this paper is deriving a basis of tensor structures in CFT and, as such, the normalization convention does not play an important role.}
A scalar three-point function can be computed by integrating over the bulk-to-boundary propagators as
\begin{align}
    \LA \phi_{\Delta_1}\phi_{\Delta_2}\phi_{\Delta_3}\RA &= \int_{\text{AdS}_{d+1}} dY\, \Pi_{\Delta_1}(X_1,Y)\Pi_{\Delta_2}(X_2,Y)\Pi_{\Delta_3}(X_3,Y)\,,
\end{align}
where $\phi_{\Delta_i}$ denotes a scalar with scaling dimension $\Delta_i$. We use the symbol $\phi$ instead of $\O$ to indicate that these scalars are not necessarily operators that are present in a CFT. 
Evaluating the integral gives~\cite{Freedman:1998tz}
\begin{align}
	\LA\phi_{\Delta_1}\phi_{\Delta_2}\phi_{\Delta_3}\RA &=\frac{b_{\Delta_1\Delta_2\Delta_3}}{X_{12}^{\Delta_{123}}X_{23}^{\Delta_{231}}X_{31}^{\Delta_{312}}}\,,\label{seed}\\[3pt]
    b_{\Delta_1\Delta_2\Delta_3} &= N_{\Delta_1}N_{\Delta_2}N_{\Delta_3}\,\frac{\pi^{\frac{d}{2}}\Gamma(\tfrac{\Delta_{123}}{2})\Gamma(\tfrac{\Delta_{231}}{2})\Gamma(\tfrac{\Delta_{312}}{2})\Gamma(\tfrac{\Delta_1+\Delta_2+\Delta_3-d}{2})}{2\Gamma(\Delta_1)\Gamma(\Delta_2)\Gamma(\Delta_3)}\,,
\end{align}
where $X_{ij} \equiv -2X_i\cdot X_j$ and $\Delta_{abc} \equiv \Delta_a+\Delta_b-\Delta_c$. 

Similarly, spinning three-point functions in AdS can be computed using spinning bulk-to-boundary propagators (see \cite{Costa:2014kfa} and Appendix~\ref{app:bulk}).
Given that the number of independent correlators is identical between AdS$_{d+1}$ and a generic CFT$_d$, these AdS correlators (or their linear combinations) can be used as a convenient kinematical basis for three-point tensor structures in CFT.
Rather than explicitly doing the bulk integrals, there exists a more efficient method to compute these spinning correlators entirely on the boundary side, effectively recycling the computation of the scalar three-point function.
This makes use of certain differential operators called weight-shifting operators, which we discuss next.

\subsection{Weight-Shifting Operators}\label{sec:WS}

Weight-shifting operators~\cite{Karateev:2017jgd} provide an efficient means to generate conformal structures with nonzero spin from scalar objects. 
These operators can be constructed by first writing down general conformally-covariant objects with the desired homogeneity in $X$ and $Z$, and then demanding that they have a well-defined projection onto the Euclidean section of the lightcone. A generic weight-shifting operator can be denoted as
\begin{align}
\cD_{\alpha \beta}^{A_1A_2\cdots} : [\Delta,\ell] \mapsto [\Delta+\alpha,\ell+\beta]\,,
\end{align}
where the superscript carries embedding-space indices, while $\alpha$ and $\beta$ in the subscript indicate the shift in the scaling dimension and spin of a conformal primary on which the operator acts, respectively. 
While there exist, in principle, infinitely many such weight-shifting operators, we will present below some particularly useful ones that will be used in our analysis. 

\paragraph{Weight-shifting operators}

Weight-shifting operators are first constructed as conformally-covariant differential operators that transform in finite-dimensional representations of the conformal group SO($d+1,1$). These can be viewed as efficient kinematical tools for generating the structures we want.
The most useful ones for our purposes arise from the vector and adjoint representations, which we classify below.

\noindent{\it Vector representation.}---We first summarize the weight-shifting operators in the vector representation of the conformal group. These are operators that carry a single embedding-space index, some of which are given by~\cite{Karateev:2017jgd, Costa:2018mcg}
\begin{align}
	\cD_{-0}^A &= X^A \,,\\
\cD_{0+}^A &= \Big((\Delta+\ell) \delta_B^A + X^A \partial_{X^B}\Big) Z^B \,,\\
\cD_{+0}^A &= \Big( c_1 \delta^A_B + X^A \partial_{X^B} \Big)
\Big( c_2 \delta^B_C + Z^B \partial_{Z^C} \Big)
\Big( c_3 \delta^C_D - \partial^C_{Z} Z_{D}  \Big) \partial_{X_D} \,,
\end{align}
where $\delta^A_B$ is the Kronecker delta and
\beq
c_1 = 2 - d + 2 \Delta\,, \quad
c_2 = 2 - d + \Delta - \ell\,, \quad
c_3 = \Delta + \ell \,.
\eeq
The above three differential operators lowers the scaling dimension, raises the spin, and raises the scaling dimension, respectively, of a conformal primary by one unit. 
Hereafter, the numbers $\ell= Z\cdot \partial_{Z}$ and $\Delta= -X\cdot\partial_{X}$ that appear on the right-hand side of these expressions always denote the spin and dimension of a conformal primary $\O_{\Delta,\ell}$ \textit{before} acting with the differential operator.

Natural choices for the normalization of the above weight-shifting operators were introduced in~\cite{Lee:2022fgr, Li:2022tby} based on their behavior in the flat-space limit. We follow \cite{Lee:2022fgr} and define the following normalized version of the operators:
\begin{align}
	\cX^A &\equiv -i \cD_{-0}^A \,,\label{cX}\\
	\cE^A &\equiv \frac{1}{\Delta-1+\ell} \cD_{0+}^A \label{cE}\,,\\
	\cP^A &\equiv \frac{2i}{(\ell+1)(\Delta-1+\ell)(d-\Delta-2+\ell)} \cD_{+0}^A \,.\label{cP}
\end{align}
Their normalization (with factors of $i$) is chosen such that the action of these operators on the scalar seed \eqref{seed} has a natural meaning in momentum space, when compared with on-shell amplitudes in the flat-space limit.\footnote{To go to momentum space, we first contract the embedding-space indices, project onto the physical position space $\mathbb{R}^d$, and then take the Fourier transform.}

To gain a more physical intuition for these weight-shifting operators, it is instructive to look at their action on bulk-to-boundary propagators. 
This allows us to derive differential representations for spinning correlators straight from the bulk integrals, as demonstrated in \cite{Li:2022tby} and Appendix~\ref{app:bulk}. 
Here, let us simply highlight an interesting ``shadow relation" between the two operators $\cP^A$ and $\cX^A$ when acting on scalar propagators:
\begin{align}\label{PXshadow}
	\cP^A\Pi_{\Delta-1}-\cX^A\Pi_{\Delta+1} = i(\tDelta-\Delta)Y^A\Pi_\Delta\,.
\end{align}
From this relation, it follows that switching between two differential representations involving $\cP^A$ and $\cX^A$ amounts to simply shifting seed dimensions and exchanging $\tDelta$ with $\Delta$! 
We will demonstrate the utility of this relation in Section~\ref{sec:ampbasis} and provide more details in Appendix~\ref{app:bulk}.

\noindent{\it Adjoint representation.}---Another useful class of weight-shifting operators are those that belong to the adjoint (antisymmetric tensor) representation of the conformal group.
We find the following two operators in the adjoint representation to be particularly useful:\footnote{Our convention for (anti)symmetrization is $T_{[A, B]}=T_{AB}-T_{BA}$ and $T_{(AB)} = T_{AB}+T_{BA}$.}
\begin{align}
	\cD_{00}^{AB} &=  \frac{1}{\sqrt 2}(X_{\phantom{X}}^{[A}\partial_X^{B]}+Z_{\phantom{X}}^{[A}\partial_Z^{B]}) \equiv \cL^{AB}\,,\label{D00}\\
	\cD_{-+}^{AB} &= \frac{1}{\sqrt 2} X^{[A} Z^{B]}\,.\label{Dmp}
\end{align}
The first operator is the usual conformal generator $\cL^{AB}$, which does not change any quantum numbers; the second operator, on the other hand, raises the spin and lowers the dimension of a conformal primary by one unit each. We define the normalized version of the latter as
\begin{align}
    \cH^{AB} \equiv \frac{i}{\Delta -2}\cD_{-+}^{AB}\,.
\end{align}
Note that this is just an algebraic factor, so its action is rather simple.

\paragraph{Bi-local operators}
The weight-shifting operators in the vector and adjoint representations that we have introduced above are conformally \textit{covariant}, carrying free embedding-space indices.  
However, it is often beneficial to have operators that are conformally \textit{invariant}. 
Such operators are especially useful if we want to project them onto position space or take their Fourier transform. 
An easy way to generate these conformally-invariant operators is by simply pairing the covariant objects into SO($d+1,1$) singlets that are {\it bi-local}, acting on two different conformal primaries.
We present below some useful bi-local operators that we will use later.

\noindent{\it Vector representation.}---Let us first consider combining the weight-shifting operators \eqref{cX}--\eqref{cP} from the vector representation. 
We can pair these operators in three different ways to form the following spin-raising combinations:
\begin{align}
	\EE{ij} &\equiv \EEc{i}{j}\,,\label{cEE}\\
 \DD{ij} &\equiv \DDc{i}{j}\,,\label{cDt}\\ \DDt{ij} &\equiv \DDtc{i}{j}\,,\label{cD}
\end{align}
where $i,j$ denote the positions that the operators act on. 
Specifically, they have the following action: $\EE{ij}$ raises the spin at both positions $i$ and $j$ by one unit; $\DD{ij}$ raises the spin at position $i$ and lowers the dimension at position $j$ by one unit each; $\DDt{ij}$ raises the spin at position $i$ and raises the dimension at position $j$ by one unit each.
We can also form the following bi-local operators that only change the scaling dimensions: 
\begin{align}
\cX_{ij}&\equiv \cX_i\cdot\cX_j\,,\label{cXX}\\ \cP_{ij}&\equiv \cP_i\cdot\cP_j\,,\label{cPP}
\end{align}
where $\cX_{ij}$ ($\cP_{ij}$) lowers (raises) the scaling dimensions by one unit at both positions $i$ and $j$. 
A summary of these bi-local operators is provided in Table~\ref{tab:bilocal}.

\SetTblrInner{rowsep=0pt}
\begin{table}[t!]
\begin{center}
\begin{tblr}{|c | c |[dotted] c| c |[dotted] c| c |}
\hline
Operator & $\Delta_i$ & $\Delta_j$ & $\ell_i$ & $\ell_j$   & Ref. \\
\hline 
$\EE{ij}$  &  0 & 0& +1& +1 & \eqref{cEE} \\ 
$\cH_{ij}$  &  $-1$ & $-1$ & +1& +1 & \eqref{cHH} \\\hline[dotted]
$\DD{ij}$ &   0 & $-1$ & $+1$& 0 & \eqref{cDt}\\
$\DDt{ij}$ &   0 & $+1$ & $+1$& 0 & \eqref{cD}\\
$\cF_{ij}$  &$-1$ & 0 & +1 & 0 & \eqref{cF} \\ \hline[dotted]
$\XX{ij}$ & $-1$ & $-1$ & 0 & 0 & \eqref{cXX} \\
$\PP{ij}$ & $+1$ & $+1$ & 0 & 0  & \eqref{cPP}\\
\hline
\end{tblr}
\end{center}
\vspace{-0.5cm}
	\caption{A selection of bi-local operators and their shift in scaling dimension and spin.}
	\label{tab:bilocal}
\end{table}

\noindent{\it Adjoint representation.}---From the weight-shifting operators \eqref{D00} and \eqref{Dmp} in the adjoint representation, the two spin-raising combinations we can take are 
\begin{align}
  \cF_{ij} & \equiv i\cH_i\cdot  \cL_{j}\,, \label{cF}\\
\cH_{ij} &\equiv \cH_i\cdot  \cH_j =\frac{\hat H_{ij}}{(\Delta_i-2)(\Delta_j-2)}\,,\label{cHH}
\end{align}
where $\cdot$ means the component-wise product in this case. The action of these operators is as follows: $\cF_{ij}$ raises the spin and lowers the dimension by one unit at the same position~$i$; $\cH_{ij}$ raises the spins and lowers the dimensions at both position $i$ and $j$ by one unit.  
For later convenience, we also defined $\hat H_{ij}\equiv -\frac{1}{2}X_i^{[A} Z_i^{B]}X_{j [A} Z_{jB]} = \frac{1}{2}H_{ij}$, which differs from the standard definition $H_{ij}$ by a factor of $\frac{1}{2}$. 
See Table~\ref{tab:bilocal} for a summary of these operators.

Notice that $\cH_{ij}$ defined above is purely algebraic. As we will see later, this feature is advantageous when it comes to an explicit evaluation of spinning correlators. 
In addition, there is another spin-raising, algebraic invariant that we can form by combining weight-shifting operators in different representations, given by 
\begin{align}
    \cV_{i,jk}  &\equiv 
    \frac{1}{2} \frac{ \Delta_j + \Delta_k - \Delta_i+\ell_i} {\Delta_i - 2} iV_{i,jk}\,,
    \label{eq:Va1}
\end{align}
where we have related it to the standard definition $V_{i,jk}= -  \frac{X_i^{[A} Z_i^{B]}X_{j,A}X_{k,B}}{X_{j} \cdot X_k}$. 
This raises the spin and lowers the dimension by one unit at the same position $i$, and it is antisymmetric in $j$ and $k$.
Its $\ell$-th power can be expressed as
\begin{align}
    \cV_{i,jk}^\ell = \frac{ (\frac{\Delta_j+\Delta_k-\Delta_i}{2})_\ell}{(\Delta_i-\ell-1)_\ell}\, (iV_{i,jk})^\ell\,,
\end{align}
where $(\cdot)_\ell$ denotes the Pochhammer symbol.
The above normalization is chosen such that we can freely replace the differential operators $\cD_{ij}$ or $\wt\cD_{ij}$ with the algebraic $\cV_{i,jk}$ when acting on a scalar seed; for example,
\begin{align}
\cV_3^\ell \LA \phi_{\Delta_1} \phi_{\Delta_2} \phi_{\Delta_3+\ell} \RA &=  \DDt{31}^\ell \LA \phi_{\Delta_1-\ell} \phi_{\Delta_2} \phi_{\Delta_3} \RA= (-1)^\ell\DDt{32}^\ell \LA \phi_{\Delta_1} \phi_{\Delta_2-\ell} \phi_{\Delta_3} \RA\nn
&=  \DD{31}^\ell \LA \phi_{\Delta_1+\ell} \phi_{\Delta_2} \phi_{\Delta_3} \RA= (-1)^\ell\DD{32}^\ell \LA \phi_{\Delta_1} \phi_{\Delta_2+\ell} \phi_{\Delta_3} \RA\, ,
\end{align}
where we define $\cV_i \equiv \cV_{i,jk}$ with cyclic ordering.

\subsection{From Amplitudes to Correlators}\label{sec:amptocorr}

Up to this point, we have introduced the ingredients for constructing three-point tensor structures in CFT, namely the scalar seeds and weight-shifting operators. 
We now describe a prescription for constructing differential representations for a basis of conformally-invariant tensor structures, mirroring the classification of scattering amplitudes.

In the conventional method of deriving differential representations for conformal correlators, one begins with a general ansatz consisting of differential operators, and then determines their coefficients by imposing physical constraints like conservation. 
Although this can be straightforwardly done case by case, the differential operators involved are generally non-commutative and can be normalized in arbitrary ways, and the same result can be obtained via many different weight-shifting paths.
In this context, the differential operators used are effectively just tools for getting the desired answer, lacking an inherent physical meaning themselves.

Recently,~\cite{Lee:2022fgr,Li:2022tby} highlighted that certain weight-shifting operators, when appropriately normalized and ordered, have a natural interpretation as being the AdS analogs of on-shell kinematic building blocks for scattering amplitudes (see also~\cite{Arkani-Hamed:2018kmz,Baumann:2019oyu,Baumann:2020dch,Diwakar:2021juk,Cheung:2022pdk, Herderschee:2022ntr} for related studies of differential representations for (A)dS correlators).
To see this, first, individual weight-shifting operators are normalized such that, in the flat-space limit, their action on scalar seeds turns into the corresponding amplitude components without extra numerical factors.
This ensures that the basic building blocks for correlators combine in the same manner as they do for amplitudes.
Moreover, while the amplitude building blocks are algebraic, the associated differential operators in AdS have an intrinsic ordering ambiguity.
To establish a well-defined mapping between these building blocks, a specific \textit{normal ordering} prescription is required for the differential operators.

\begin{figure}[t!]
\begin{center}
\begin{tikzpicture}[baseline=(current  bounding  box.center)]
\node at (-4,0)  {Amplitudes in $\mathbb{R}^{1,d}$};
\node at (0,0.4)  {\small uplifting};
\node at (0,-0.4)  {\small flat-space limit};
\draw [thick,-stealth] (-1.5,0.1) -- (1.5,0.1);
\draw [thick,-stealth] (1.5,-0.1) -- (-1.5,-0.1);
\node at (4.2,0)  {Correlators in AdS$_{d+1}$};
\end{tikzpicture}
\captionof{figure}{Relation between scattering amplitudes and AdS correlators.}
\label{fig:comm}
\end{center}
\end{figure}
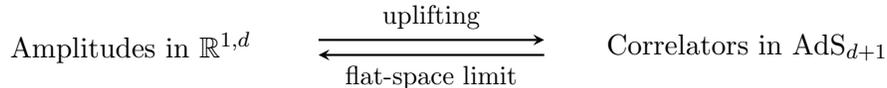

It is well known that scattering amplitudes can be obtained from AdS correlators by taking the flat-space limit.
At the same time, understanding the precise mapping between the algebraic and differential building blocks in the two spacetimes also enables us to ``uplift'' flat-space amplitudes to AdS (or dS) through a direct substitution of the basis elements (see Figure~\ref{fig:comm}). 
Given that a conformal three-point function is non-perturbatively determined up to a few constants, the AdS correlators obtained through this method can serve as a basis of three-point structures for general CFTs. 

In order to generate the desired tensor structures for conserved spinning correlators, we need to combine multiple differential operators that act on scalar seeds with different scaling dimensions. 
We have a set of two operators that we can use for this purpose, $\cP^A$ and $\cX^A$, which either raises or lowers scaling dimensions.
Each choice of these operators offers its own set of benefits. We will outline the basic idea below, and demonstrate the procedure with concrete examples in Section~\ref{sec:ampbasis}.

\paragraph{$\cP$ representation}
One way of constructing differential representations for AdS correlators involves using $\cP^A$ to shift the seed dimensions. 
In this approach, we take a given scattering amplitude that depends on a polarization vector $\eps^\mu$ and momentum $p^\mu$ (see \S\ref{sec:amp}), and make the following replacement:
    \begin{align}\label{replace}
	\epsilon^\mu \to \cE^A \,,\quad p^\mu \to \cP^A \, .
\end{align}
Despite the embedding-space index $A$ having a larger range than the spacetime index $\mu$, the above replacement is always applied to contracted vectors; e.g.~$ \epsilon^\mu p_\mu \rightarrow \cE^A \cP_A$. 
The operators are then ordered such that all $\cE^A$ are placed to the {\it left} of $\cP^A$.  As shown in~\cite{Li:2022tby} and Appendix~\ref{app:bulk}, this operator ordering follows from certain identities that re-express bulk covariant derivatives acting on a spinning propagator as normal-ordered $\cE^A$ and $\cP^A$ operators acting on scalar propagators.

\newpage
A key advantage of this ``$\cP$ representation'' is its simplicity in verifying the conservation condition~\cite{Li:2022tby}. 
To see this, let us define the divergence operator as
\begin{align}
	\text{div}  &\equiv  2i \,\partial_{X} \cdot T_{Z} \, ,\label{div}\\
T_Z^A &\equiv  \bigg(\frac{d}{2} - 1 + Z \cdot \partial_{Z} \bigg) \partial_{Z}^A - \frac{1}{2} Z^A \partial_{Z} \cdot \partial_{Z}\,,\label{eq:Todorov}
\end{align}
where $T_Z^A$ is the Todorov operator that strips off a null vector $Z^A$~\cite{Dobrev:1975ru}.\footnote{Note that $X^A$ is null, and the operator $\cP^A$ simplifies to $\cP^A=-2iT_X^A$ when acting on a scalar correlator.}  Taking the divergence of a conserved tensor should then annihilate its correlator (modulo contact terms).
Our choice of the normalization in \eqref{cE}, \eqref{cP}, and \eqref{div} implies that
\beq
\begin{aligned}\label{divE}
	\text{div}\, \cE^A &= \cP^A\,,\\ \text{div}\, \cE^A\cE^B &= 2\cP^A\cE^B\,,
\end{aligned}
\eeq
when acting on a conserved spin-1 current and a stress tensor, respectively. 
Remarkably, this is analogous to the replacement $\eps^\mu\to p^\mu$ under which flat-space amplitudes of massless particles vanish due to gauge invariance. 
This similarity suggests that one can directly uplift a given scattering amplitude of photons or gravitons to AdS. 
While this is not entirely straightforward in all cases because the operators on the right-hand side of \eqref{divE} are no longer normal ordered, we will show in Section~\ref{sec:ampbasis} that this approach is indeed valid for the three-point structures that we study in this paper.

Despite this conceptual advantage, one drawback of this approach is the necessity to evaluate multiple normal-ordered operators $\cE^A$ and $\cP^A$ before contracting their indices, resulting in an increased computational complexity. 
For example, consider the expansion of the following normal-ordered operators:
\begin{align}
    :\!\DDt{12}\DDt{21}\!:\, 
    &=\cE_1^A \cE_2^B \cP_{2A}\cP_{1B}\,, \label{DDid}\\
    :\!\DDt{12}^2\DDt{21}^2\!:\, 
    &=\cE_1^{A_1}\cE_1^{A_2} \cE_2^{B_1}\cE_2^{B_2} \cP_{2A_1}\cP_{2A_2}\cP_{1B_1}\cP_{1B_2}\,,\label{DDid2}
\end{align}
where the colons on the left-hand side indicate normal ordering, and the operators on the right-hand side must act on scalar seeds before all of their indices are contracted. 
This proliferation of indices becomes increasingly difficult to handle for higher spins.

\paragraph{$\cX$ representation}

As an alternative approach, we can instead make the following replacement of the amplitude building blocks:
    \begin{align}\label{replace}
	\epsilon^\mu \to \cE^A \,,\quad p^\mu \to \cX^A \, .
\end{align}
Interestingly, the shadow relation~\eqref{PXshadow} implies that this uplifting procedure applies directly, not to conserved tensors themselves, but instead to their shadows.
In order to obtain correlators of conserved tensors, we can follow a straightforward procedure to translate a given $\cP$ representation to its $\cX$ counterpart, as we outline below.

As an initial move, after having obtained the $\cP$ representation of a correlator, we simply replace the operators $\cP^A\to \cX^A$. 
Interestingly, this move naturally leads to the emergence of bi-local operators and algebraic invariants. 
For example, we have the following interesting identities involving normal-ordered operators:
    \begin{align}\label{DDid3}
        :\!\DD{12}\DD{21}\!:\, &= \EE{12}\XX{12}-\hat H_{12}\,,\\
        :\!\DD{12}^2\DD{21}^2\!:\, &= \EE{12}^2\XX{12}^2-2\hat H_{12}\EE{12}\XX{12}+\hat H_{12}^2\,.\label{DDid4}
    \end{align}
Notice that the right-hand side simply involves products of bi-local operators and the algebraic factor $\hat H_{12}$; evaluating these expressions is much more efficient than those in terms of $\cP^A$, given in ~\eqref{DDid} and \eqref{DDid2}. 
Moreover, this ``$\cX$ representation'' is also useful for momentum-space calculations, since it is straightforward to Fourier transform bi-local operators~\cite{Baumann:2019oyu,Baumann:2020dch}. 
However, in this approach we sometimes need to incorporate extra terms beyond those obtained from a direct uplifting procedure, in order to satisfy the conservation condition. See Section~\ref{sec:ampbasis} for concrete applications of this approach.

To summarize, $\cP$ representations make verifying the conservation constraint straightforward, but they are often infeasible for explicit evaluations.
In contrast, deriving $\cX$ representations is relatively less straightforward, but they are computationally much more efficient.
In the next section, whenever appropriate, we will construct spinning three-point structures using both approaches.

\section{Three-Point Functions in Amplitude Basis}\label{sec:ampbasis}

In the previous section, we outlined the ingredients of the differential representations of conformal correlators and described the uplifting procedure. 
We now use this methodology to systematically construct a range of spinning three-point structures involving conserved currents or stress tensors and non-conserved spinning operators.

\subsection{Amplitude Basis}\label{sec:3pt}

We begin by reviewing the basics of three-particle scattering amplitudes (see e.g.~\cite{Costa:2011mg,Bonifacio:2017nnt,Afkhami-Jeddi:2018apj}) and three-point functions in CFT. We then define a basis for conformally-invariant three-point tensor structures based on the classification of scattering amplitudes.

\subsubsection{Three-Particle Amplitudes}\label{sec:amp}

Consider the scattering of three particles with arbitrary spins in $D=d+1$ spacetime dimensions, where each particle carries the momentum $p_i^\mu$ and polarization tensor $\eps_i^{\mu_1\cdots\mu_\ell}$. 
The polarization tensors are symmetric, traceless, and transverse, and it is convenient to express them as a product of auxiliary vectors $\eps_i^\mu$ with the replacement $\eps_i^{\mu_1\cdots\mu_\ell} \to  \eps_i^{\mu_1}\cdots \eps_i^{\mu_\ell}$, which are transverse ($p_i\cdot \eps_i=0$) and null ($\eps_i^2=0$). 

A three-particle amplitude is a Lorentz-invariant function of $p_i^\mu$ and $\eps_i^\mu$, and takes the general form
\begin{align}
\cA_{\ell_1\ell_2\ell_3} = \sum_{\mathbf{n},\mathbf{m}} b_{\mathbf{n},\mathbf{m}}\,\epsilon_{12}^{n_{12}} \epsilon_{23}^{n_{23}} \epsilon_{31}^{n_{31}} d_{12}^{m_{1}} d_{23}^{m_{2}}  d_{31}^{m_{3}}\, \delta^{(d+1)}(p_1+p_2+p_3)\,,\label{eq:genamp}
\end {align}
where $b_{\mathbf{n},\mathbf{m}}$ are numerical coefficients, and we have introduced the following shorthand notation
\begin{equation}
    \ee{ij} \equiv \eec{i}{j}\,,\quad \dd{ij}\equiv \ddc{i}{j}\,.
\end{equation}
We have dropped $p_{ij}\equiv p_i\cdot p_j$ here, which for three-particle scattering can be written in terms of particle masses using the on-shell condition and momentum conservation.
The amplitude is a homogeneous polynomial of degree $\ell_i$ in each $\eps_i$, which means that the integer parameters $\mathbf{n}\equiv\{n_{12},n_{23},n_{31}\}$ and $\mathbf{m}\equiv\{m_{1},m_{2},m_{3}\}$ satisfy the constraints
\beq
\begin{aligned}\label{nmcond}
    m_i + \sum_{j\ne i} n_{ij}  = \ell_i \,,
\end{aligned}
\eeq
for $i=1,2,3$. These have a finite number of solutions for general spins, determined by a simple counting formula~\cite{Costa:2011mg}.\footnote{The case of $D=d+1=4$ spacetime dimensions needs to be treated separately, as five four-vectors cannot all be linearly independent. 
This linear dependence is captured by the Gram determinant involving the five vectors $\epsilon_1,\epsilon_2,\epsilon_3,p_1,p_2$, which can lead to a reduction of the number of independent amplitudes.} 
If the particle $i$ is massless, then the amplitude must be invariant under the gauge transformation $\epsilon_i^\mu \to \epsilon_i^\mu + \xi p_i^\mu$ and satisfy the on-shell Ward identity $p_{i,\mu} \cA_{\ell_1\ell_2\ell_3}^\mu=0$. 
In general, imposing this constraint reduces the number of independent amplitudes.

\subsubsection{Three-Point Functions}

Consider now spinning three-point functions in CFT$_d$.
We will use an index-free notation and denote a spin-$\ell$ operator in embedding space as $\O_\ell = Z^{A_1}\cdots Z^{A_\ell}\O_{A_1\cdots A_\ell}$. 
Similar to what is done for amplitudes, we can express the general three-point function in terms of the algebraic building blocks $H_{ij}$ and $V_i$ as~\cite{Costa:2011mg}
\begin{align}
    \LA \O_{\ell_1}\O_{\ell_2}\O_{\ell_3}\RA = \sum_{\mathbf{n} ,\mathbf{m}} c_{\mathbf{n} ,\mathbf{m}}  \frac{H_{12}^{n_{12}} H_{13}^{n_{13}} H_{23}^{n_{23}} V_1^{m_{1}} V_2^{m_{2}}V_3^{m_{3}}}{X_{12}^{\tau_{123}}X_{23}^{\tau_{231}}X_{31}^{\tau_{312}}}\,,\label{HV3pt}
\end{align}
where $c_{\mathbf{n} ,\mathbf{m}}$ are numerical coefficients, $\tau_{ijk}\equiv\tau_i+\tau_j-\tau_k$, and $\tau_i=\Delta_i-\ell_i$ is the twist. 
The integer parameters $n_{ij}$ and $m_i$ obey the same constraints \eqref{nmcond} as before, so that the number of independent correlators is exactly the same as that of independent amplitudes for a given set of spins.  
If the scaling dimension of $\O_{\ell_i}$ saturates the unitarity bound, $\Delta_i=d-2+\ell_i$, then the operator is conserved and the associated divergence of the correlator must vanish, $\text{div}_i\LA\O_{\ell_1}\O_{\ell_2}\O_{\ell_3}\RA=\text{``0,''}$ up to possible contact terms that vanish at separated points.
The conservation constraint is the analog of the on-shell Ward identity for amplitudes of massless particles, and imposing this generally reduces the number of independent structures.

Conformal three-point functions with more than one spinning operator generally depend on more than one OPE coefficient, so it is useful to have an organizing principle to systematically classify their independent tensor structures.
In the case of three-particle amplitudes~\eqref{eq:genamp}, an organization based on distinct values of $m_i$ is useful, as it makes clear the power counting of the effective field theory that gives these amplitudes.
In CFTs, one can similarly organize OPE coefficients by inverse powers of $\Delta_{\rm gap}\gg 1$, where $\Delta_{\rm gap}$ is the characteristic scale associated with higher-spin operators with spin greater than two.
This organization is particularly useful in the holographic context, where large-$N$ theories with a large higher-spin gap are conjectured to admit a local gravity dual in AdS~\cite{Heemskerk:2009pn}. 
In such scenarios, the inverse powers of $\Delta_{\rm gap}$ are expected to correspond to the counting of derivatives in AdS~\cite{Heemskerk:2009pn, Camanho:2014apa, Afkhami-Jeddi:2016ntf,Caron-Huot:2017vep,Kulaxizi:2017ixa, Costa:2017twz, Meltzer:2017rtf,Belin:2019mnx,Kologlu:2019bco,Caron-Huot:2021enk}.

As an illustrative example, consider the three-point structures from the Yang-Mills and $F_{\mu\nu}^3$ cubic interactions, with $F_{\mu\nu}$ being the field-strength tensor. 
The on-shell amplitudes are given by
\begin {align}
\cA_{\rm YM} & \propto \epsilon_{12}d_{31} +  \text{cyc.}\,, \label{eq:AYM} \\
\cA_{F^3} & \propto d_{12}d_{23}d_{31} \, , \label{eq:AF3}
\end {align}
where we have suppressed the color indices and coupling constants.
These two amplitudes are clearly distinguished by their mass dimensions. 
On the other hand, the corresponding AdS boundary three-point functions, expressed in the algebraic basis, take the form
\begin{align}
    \LA JJJ\RA_{\rm YM} &\propto \frac{(H_{23}V_1+\text{cyc.})+\frac{3(d-2)}{2d-3}V_1V_2V_3}{(X_{12}X_{23}X_{31})^{d/2}}\,,\\
    \LA JJJ\RA_{F^3} &\propto \frac{(H_{23}V_1+\text{cyc.})+(d+2)V_1V_2V_3}{(X_{12}X_{23}X_{31})^{d/2}}\,.
\end{align}
We see that the counting of $\Delta_{\rm gap}$ is not manifest in the algebraic basis, with the two correlators just differing in the coefficient of $V_1V_2V_3$.

An alternative idea of organizing three-point structures in CFT is to choose a basis that clearly reveals the amplitude structures in the flat-space limit.
As discussed in the previous section, this can be naturally accomplished with canonically-normalized weight-shifting operators. 
From the amplitudes \eqref{eq:AYM} and \eqref{eq:AF3}, the uplifting procedure $\eqref{replace}$ gives
\begin{align}
\LA JJJ\RA_{\rm YM} & =\EE{12}\DDt{31}\LA\phi_{d-2}\phi_{d-1}\phi_{d-1}\RA+\text{cyc.}\,, \label{eq:JJJYMup} \\
\LA JJJ\RA_{ F^3} & = \ :\!\DDt{12}\DDt{23}\DDt{31}\!: \LA\phi_{d-2}\phi_{d-2}\phi_{d-2}\RA\,, \label{eq:JJJF3up}
\end{align}
where the differential operators act on scalar seed three-point functions.
We can think of the scalar seeds as playing an identical role as the momentum-conserving delta function in \eqref{eq:genamp}. 
The operators in \eqref{eq:JJJF3up} are normal ordered and are given by
\begin{align}
:\!\DDt{12}\DDt{23}\DDt{31}\!:\ \equiv \ :\!(\DDtc{1}{2})(\DDtc{2}{3}) (\DDtc{3}{1}):\ =\cE_1^A\cE_2^B \cE_3^C \cP_{2A}\cP_{3B} \cP_{1C}\,.
\end{align}
The counting of $\Delta_{\rm gap}$ is now transparent in the differential representations \eqref{eq:JJJYMup} and \eqref{eq:JJJF3up}, where the number of $\cP_j^A$ (or $\DDt{ij}$) is associated to the number of bulk AdS derivatives or the $\Delta_{\rm gap}$ counting in the dual CFT~\cite{Li:2022tby}. 
Specifically, $\LA JJJ\RA_{F^3}\sim \Delta_{\rm gap}^{-2}$, while the Ward identity implies that $\LA JJJ\RA_{\rm YM}\sim c_J$, where $c_J$ is the normalization of the two-point function $\LA JJ\RA$. 
Moreover, checking conservation of these expressions is now trivial with the use of the identity $\text{div}\,\cE^A = \cP^A$ shown in \eqref{divE}.
Likewise, the differential representation for $\LA TTT\RA$ clearly exhibits the double-copy relation with $\LA JJJ\RA$ and thus also the counting of $\Delta_{\rm gap}$~\cite{Lee:2022fgr, Li:2022tby}.

The correlators $\LA JJJ\RA_{\rm YM}$ and $\LA JJJ\RA_{F^3}$, despite having a bulk origin, span the two-dimensional vector space of independent tensor structures for $\langle JJJ \rangle$ in any CFT. 
We will call such basis of tensor structures that arise from the uplifting of amplitudes as an \textit{amplitude basis}.
For simplicity, when the context is clear we will not explicitly show scalar seeds, but instead simply show the differential part of correlators as
\begin{align}
\LA\wh{JJJ}\RA_{\rm YM} & =\EE{12}\DDt{31}+\EE{23}\DDt{12}+\EE{31}\DDt{23} \,, \label{eq:MYM}\\
\LA \wh{JJJ}\RA_{F^3} & = \ :\!\DDt{12}\DDt{23}\DDt{31}\!: \label{eq:MF3}\, ,
\end{align}
where $\LA\wh{\cdots}\RA$ means that we are stripping off the scalar seed functions, with the understanding that individual terms act on seeds with possibly different scaling dimensions.

Our goal is to extend this approach to three-point functions involving both conserved and non-conserved spinning operators. 
This requires first classifying flat-space amplitudes and then uplifting these expressions to AdS. 
The resulting differential representations then serve as a basis for three-point tensor structures in CFT.

\subsection{$\langle JJ\mathcal{O}_\ell \rangle$ and $\langle TT\mathcal{O}_\ell \rangle$}

In this subsection, we present differential representations for three-point structures involving two conserved operators and one non-conserved, symmetric spinning operator.

\subsubsection{$\langle JJ\mathcal{O}_\ell \rangle$}

Let us first summarize the amplitude structures. In $D=d+1\ge 8$ spacetime dimensions, there exist six independent parity-even amplitudes of two photons and one massive spinning particle, which can be classified according to the irreducible representations of the massive little group SO($d$)~\cite{Chakraborty:2020rxf}. 
If we restrict to totally symmetric representations in $D\ge 4$, then the number of independent amplitudes reduces down to two.
These are given by
\begin{align}
\cA_{11\ell}^{(1)} &=  (\ee{12}\pp{12}-\dd{12}\dd{21})d_{31}^{\ell}\quad (\ell\ge 0)\,,\label{eq:JJOA1genl}\\
\cA_{11\ell}^{(2)} &=  (\ee{13}\ee{23}\pp{12}+\ee{12}\dd{31}\dd{32}-\ee{13}\dd{21}\dd{32}-\ee{23}\dd{12}\dd{31} )\dd{31}^{\ell-2}\quad (\ell\ge 2)\,,
\label{eq:JJOA2genl}
\end{align}
for even $\ell$. Since we are mainly interested in kinematical structures in this work, we will henceforth drop overall numerical factors in these definitions. Notice that the second amplitude only exists for $\ell\ge 2$. 
These amplitudes can be constructed as linear combinations of the independent solutions to the constraints~\eqref{nmcond}, in addition to satisfying the on-shell gauge invariance and permutation symmetry $1 \leftrightarrow 2$.
The corresponding cubic Lagrangians involving two $F_{\mu \nu}$ and one massive spin-$\ell$ field $S^{\mu_1 \cdots \mu_\ell}$ are also shown in Table \ref{tab:JJO}.

\begin{table}
\begin {center}
\begin {tblr} { | l | l | l |  } \hline
Dim. & Lagrangian & Spin  \\ \hline
$D \geq 4$ & $ \nabla_{\mu_1} \cdots \nabla_{\mu_\ell} F_{\mu \nu} F^{\mu \nu} S^{\mu_1 \cdots \mu_\ell}$  & $\ell \geq 0$, $\ell\in 2\mathbb{Z}$  \\ \hline[dotted]
$D \geq 4$  & $\nabla_{\mu_3} \cdots \nabla_{\mu_\ell} F_{\alpha \mu_1} F^{\alpha}_{\mu_2} S^{\mu_1  \cdots \mu_\ell}$ & $\ell \geq 2$, $\ell\in 2\mathbb{Z}$  \\ \hline
\end {tblr}
\caption{Cubic interactions for two photons and one massive spinning particle. There are two independent parity-even amplitudes in $D\ge 4$.}
\label{tab:JJO}
\end {center}
\end{table}

\paragraph{$\cP$ representation}
It is straightforward to construct a basis of tensor structures for $\LA JJ\O_\ell \RA$ by directly uplifting the amplitude expressions shown above. This gives the following differential representations for two tensor structures:
\begin{align}
    \LA\wh{JJ\O_\ell}\RA^{(1)} &= (\EE{12}\PP{12}\,-   :\!\DDt{12}\DDt{21}\!:)  \cV_3^\ell\,,\label{eq:A1genlEP}\\
    \LA\wh{JJ\O_\ell}\RA^{(2)} &= (\EE{13}\EE{23}\PP{12} + \EE{12}\DDt{31}\DDt{32} -\EE{13}\DDt{21}\DDt{32}-\EE{23}\DDt{12}\DDt{31} )\cV_3^{\ell-2}\,,\label{eq:A2genlEP}
\end{align}
where we used the fact that $\cV_3=\DDt{31}$ when acting on scalar seeds. 
By construction, these correlator structures have the correct flat-space limit.
To see that the above structures are conserved, we take the divergence at position 1 and use identity $\text{div}\, \cE^A = \cP^A $ in \eqref{divE}, which gives
\begin{align}
\text{div}_1 \LA\wh{JJ\O_\ell}\RA^{(1)} &=  (\DDt{21}\PP{12}   -  \DDt{21}\PP{12})\cV_3^\ell   = 0 \, ,\\
\text{div}_1  \LA\wh{JJ\O_\ell}\RA^{(2)} &= (\DDt{31}\EE{23}\PP{12} + \DDt{21}\DDt{31}\DDt{32} -\DDt{31}\DDt{21}\DDt{32}-\EE{23}\PP{12}\DDt{31} )\cV_3^{\ell-2}=0\,,
\end{align}
where we have used $[\cE_i, \cP_j] =0$ for $i\ne j$. 
Conservation at position 2 can be checked similarly. Since the above two structures are linearly independent, the three-point function $\LA JJ\O_\ell\RA$ in any CFT can be captured by a linear combination of $\LA JJ\O_\ell\RA^{(1)}$ and $\LA JJ\O_\ell\RA^{(2)}$.\footnote{An interesting special case is when $\mathcal{O}_\ell$ corresponds to the stress tensor, with $\ell =2$ and $\Delta_3 = d$. In this case, the two distinct structures for $\LA JJT \RA$ (see e.g.~\cite{Hartman:2016dxc}) can be related to  $\LA JJO_\ell\RA^{(1)}$ and $\LA JJO_\ell\RA^{(2)}$ by a basis rotation.}

We may also derive two different structures for $\LA JJ\O_\ell\RA$ by directly computing Witten diagrams in AdS. 
In Appendix~\ref{app:bulk}, we carry out this bulk computation for $\ell=0, 2$ and show that these Witten diagrams are given by linear combinations of the two structures shown above.

\paragraph{$\cX$ representation}

When it comes to an explicit evaluation of the correlators, representations involving $\cP^A$ can be inconvenient due to the necessity of contracting the indices of operators {\it after} acting them on scalar seeds.
It is therefore beneficial to have equivalent representations in terms of $\cX^A$. 

For this analysis, it is useful to first consider a more general form of the three-point function involving non-conserved spin-1 operators with arbitrary scaling dimension $\Delta_J\ge d-1$. 
It is easiest to illustrate the basic idea with the simple example of $\ell=0$, for which there is a unique three-point function $\LA JJ\O\RA$. 
An explicit AdS calculation of the three-point function of two massive spin-1 fields and a scalar gives
\begin{align}\label{JJOPrep}
	\LA\wh{ J_{\Delta_J}J_{\Delta_J} \cO }\RA &\propto \EE{12}\PP{12}\,- :\!\DDt{12}\DDt{21}\!:  -\, (\Delta_J-d+1)^2 \,\EE{12}\,,\nn
 &=  \EE{12}\XX{12}\,- :\!\DD{12}\DD{21}\!:  -\, (\Delta_J-1)^2\,\EE{12}\,,
\end{align}
see Appendix~\ref{app:bulk} for a derivation.
In the second line, we applied the shadow relation \eqref{PXshadow}, which amounts to exchanging $\{\cP,\Delta_J\}$ with $\{\cX,\tilde\Delta_J\}$.
Note that in the above $\cX$ representation, the term $:\!\DD{12}\DD{21}\!:\, = \cE_1^A\cE_2^B\cX_{2A}\cX_{1B}$ still requires the operators to be normal ordered before their indices are contracted. 
Using the identity \eqref{DDid3}, however, we can turn \eqref{JJOPrep} into
\begin{align}
\LA \wh{J_{\Delta_J} J_{\Delta_J} \O}\RA & = \hat H_{12} -  (\Delta_J-1)^2  \EE{12} \,,
\end{align}
thus effectively transforming the normal-ordered product of operators into the algebraic factor $\hat H_{12}$.
Setting $\Delta_J=d-1$ for the conserved case, we get
\begin{align}\label{JJObirep}
    \LA\wh{JJ\cO}\RA &\propto  \cH_{12} -\EE{12}  \,,
\end{align}
where the expression has been further simplified using the normalized operator $\cH_{ab}$.

A similar derivation follows for $\ell = 2$. In this case, we have two independent cubic interactions in AdS, and we first compute the corresponding three-point functions assuming generic scaling dimension $\Delta_J\ge d-1$. 
As we show in Appendix~\ref{app:bulk}, this leads to the following differential representations:
\begin{align}
    \LA\wh{J_{\Delta_J}J_{\Delta_J}\O_2}\RA^{(1)} &\propto \LA\wh{JJ\O_2}\RA^{(1)} -(\Delta_J-d+1) \Big[ (\Delta_J-d+3)\EE{12}\DDt{31}\DDt{32} \nn[-2pt]
    &\hskip 62pt+2\big((\Delta_J-d+1)\EE{13}\EE{23}-\EE{13}\DDt{21}\DDt{32}-\EE{23}\DDt{12}\DDt{31})\big)\Big]\,,\label{JJO2gen1}\\
 \LA\wh{J_{\Delta_J}J_{\Delta_J}\O_2}\RA^{(2)}\! &\propto  \LA\wh{JJ\O_2}\RA^{(2)} -(\Delta_J-d+1)^2\EE{13}\EE{23} \,,\label{JJO2gen2}
\end{align}
where $\LA\wh{JJ\O_2}\RA^{(1,2)}$ denote the structures \eqref{eq:A1genlEP} and \eqref{eq:A2genlEP} in the conserved case.
The corresponding $\cX$ representations then directly follow from the shadow relation \eqref{PXshadow}.

Generalizing to general spin~$\ell$ and setting $\Delta_J=d-1$, we find
\begin{align}
  \hskip -8pt\LA \wh{J J \O_\ell}\RA^{(1)}\!  &=\! \big(\hat H_{12} -(d-2+2\ell)\EE{12} -\ell(\EE{13}\DD{21}-\EE{23}\DD{12})\cV_3^{-1}\label{eq:A1genldown} \nn
  &\hskip 112pt+\ell(\ell-1)(d-2)\EE{13}\EE{23}\cV_3^{-2}\big)\cV_3^{\ell}\, ,\\
   \hskip -8pt \LA\wh{JJ\O_\ell}\RA^{(2)}\! &=\! \big(\EE{13}\EE{23}\XX{12} + \EE{12}\DD{31}\DD{32} - \EE{23}\DD{12}\DD{31}-\EE{13}\DD{21}\DD{32} -(d-2)^2\EE{13}\EE{23} \big)\cV_3^{\ell-2}.\label{eq:A2genldown}
\end{align}
These two differential representations can be used as a basis of tensor structures for $\LA J J \O_\ell\RA$ in CFT.
The presence of the term $\hat H_{12}$ again follows from the identity \eqref{DDid3}.
Although these expressions may appear somewhat complex, we emphasize that evaluating them in the $\cX$ representation, consisting solely of bi-local operators, is much simpler than in the $\cP$ representation. 
The benefits of this approach are even more pronounced when we consider three-point structures involving stress tensors, which we discuss next.

\subsubsection{$\langle TT\mathcal{O}_\ell \rangle$}

In the most general case of $D = d+1 \ge 8$ spacetime dimensions, there exist 20 parity-even scattering amplitudes of two gravitons and one massive spinning particle, which have been classified in~\cite{Chakraborty:2020rxf, Caron-Huot:2022jli}.
Restricting to totally symmetric tensor representations, the number of independent amplitudes reduces to three in $D \geq 5$ and two in $D = 4$.
These arise from high-derivative interactions involving two Riemann tensors $R_{\alpha \beta \gamma \delta}$ and one massive spin-$\ell$ field $S^{\mu_1 \cdots \mu_\ell}$, which are shown in Table~\ref{tab:TTO}. 
The corresponding on-shell amplitudes are
\begin{align}
   \hskip -10pt \cM_{22\ell}^{(1)} &=(\eps_{12}p_{12}-d_{12}d_{21})^2d_{31}^\ell\quad (\ell\ge 0)\,, \label{eq:TTOM1genl}\\
   \hskip -10pt \cM_{22\ell}^{(2)} &=(\eps_{12}p_{12}-d_{12}d_{21})(\ee{13}\ee{23}\pp{12}+\ee{12}\dd{31}\dd{32}-\ee{23}\dd{12}\dd{31}-\ee{13}\dd{21}\dd{32})d_{31}^{\ell-2}\ \ (\ell\ge 2)\,, \label{eq:TTOM2genl}\\
   \hskip -10pt \cM_{22\ell}^{(3)} &=(\ee{13}\ee{23}\pp{12}+\ee{12}\dd{31}\dd{32}-\ee{23}\dd{12}\dd{31}-\ee{13}\dd{21}\dd{32})^2d_{31}^{\ell-4}\quad (\ell\ge 4)\, . \label{eq:TTOM3genl}
\end{align}
The second amplitude $\cM_{22\ell}^{(2)}$ only exists for $D\ge 5$, while the other two exist for $D\ge 4$.
It should be noted that these spin-2 amplitudes can be obtained by taking two copies of the spin-1 amplitudes in \eqref{eq:JJOA1genl} and \eqref{eq:JJOA2genl}.

\begin{table}
\begin {center}
\begin {tblr} { | l | l | l |  } \hline
Dim. & Lagrangian & Spin \\ \hline
$D \geq 4$ & $ \nabla_{\mu_1} \cdots \nabla_{\mu_\ell} R_{\alpha \beta \gamma \delta} R^{\alpha \beta \gamma \delta} S^{\mu_1 \cdots \mu_\ell}$  & $\ell \geq 0$, $\ell\in 2\mathbb{Z}$   \\ \hline[dotted]
$D \geq 5$  & $ \nabla_{\mu_3} \cdots \nabla_{\mu_\ell} R_{\mu_1 \alpha \beta \gamma} R_{\mu_2}{}^{\alpha \beta \gamma} S^{\mu_1 \cdots \mu_\ell}$ & $\ell \geq 2$, $\ell\in 2\mathbb{Z}$   \\ \hline[dotted]
$D \geq 4$  & $ \nabla_{\mu_5} \cdots \nabla_{\mu_\ell} R_{\mu_1 \alpha \mu_2 \beta} R_{\mu_3}{}^{\alpha}{}_{\mu_4}{}^{\beta} S^{\mu_1  \cdots \mu_\ell}$ & $\ell \geq 4$, $\ell\in 2\mathbb{Z}$   \\ \hline
\end {tblr}
\caption{Cubic interactions for two gravitons and one massive spinning particle. There are three independent parity-even amplitudes in $D\ge 5$ and two in $D= 4$.}\label{tab:TTO}
\end {center}
\end{table}

\paragraph{$\cP$ representation}
In $d \geq 4$ spatial dimensions, uplifting the amplitude expressions shown above leads to the following differential representations for three-point structures involving two stress tensors and one non-conserved spinning operator:
\begin{align}
    \LA \wh{TT\O_\ell}\RA^{(1)}&=\ :\!(\EE{12}\PP{12}-\DDt{12}\DDt{21})^2\!:\cV_3^\ell\,, \label{TTO1P}\\
    \LA \wh{TT\O_\ell}\RA^{(2)}&=\ :\!(\EE{12}\PP{12}-\DDt{12}\DDt{21})(\EE{13}\EE{23}\PP{12}+\EE{12}\DDt{31}\DDt{32}-\EE{23}\DDt{12}\DDt{31}-\EE{13}\DDt{21}\DDt{32})\!:\cV_3^{\ell-2}\,,\nonumber\\
    \LA \wh{TT\O_\ell}\RA^{(3)} &=\ :\!(\EE{13}\EE{23}\PP{12}+\EE{12}\DDt{31}\DDt{32}-\EE{23}\DDt{12}\DDt{31}-\EE{13}\DDt{21}\DDt{32})^2\!:\cV_3^{\ell-4}\,,\label{TTO3P}
\end{align}
where the first and third structures form a complete basis in $d=3$.
The differential double-copy structure is also apparent in these expressions.
We remind the readers that weight-shifting operators, after being squared, must be normal ordered prior to contracting their indices. 

It is straightforward to verify that the above three structures are all conserved. 
For example, expanding out the product of normal-ordered operators in $\LA \wh{TT\O_\ell}\RA^{(1)}$, we get
\begin{align}
 :\!(\EE{12}\PP{12} - \DD{12}\DD{21})^2\!:\,
    &= \EE{12}^2\PP{12}^2-2\EE{12}:\!\DDt{12}\DDt{21}\!:\PP{12} \,+ :\!\DDt{12}^2\DDt{21}^2\!:\,\nn
    &=\EE{12}^2\PP{12}^2-2\EE{12}\,\cE_1^A\DDt{21} \cP_{2,A}\PP{12}  + \cE_1^A\cE_1^B\DDt{21}^2 \cP_{2A}\cP_{2B}\,.
\label{eq:B1l0EP}
\end{align}
Taking the divergence at position 1 using \eqref{divE} then gives
\begin{align}\label{divTTO1}
 \text{div}_1:\!(\EE{12}\PP{12} - \DD{12}\DD{21})^2\!:\,
    &= 2 ( \cE_2\cdot\cP_1 \wt\cA\, \cP_{12} - \cP_1^A \cE_2^B \wt\cA\, \cP_{2 A} \cP_{1B}  )\,,
\end{align}
where we defined $\wt\cA \equiv \EE{12} \PP{12} \,-\! :\!\!\DDt{12}\DDt{21}\!\!:\,$. Although the two terms on the right-hand side of \eqref{divTTO1} clearly cancel in the flat-space limit, their cancellation at the operator level is not obvious because $\cP$ does not commute with $\wt\cA$ in general. 
Nevertheless, an explicit computation confirms that the two terms do cancel when applied to a scalar seed of scaling dimension $\Delta = d-1$ at positions 1 and 2. 
The conservation of $\LA TT\O_\ell\RA^{(2,3)}$ can be verified similarly. In these cases, while more operators are required to cancel, showing this essentially boils down to proving identities similar to \eqref{divTTO1}, e.g., with $\wt\cA$ instead taking the form $\cE_{12}\DDt{31}-\cE_{13}\DDt{21}$.

As in the $\LA JJ\cO_\ell \rangle$ case, the expressions in the $\cP$ representation shown above are useful for verifying conservation. However, they are inefficient for explicit computations, requiring several differential operators to be applied before their indices can be contracted.
To this end, we now present equivalent expressions for $\LA \wh{TT\O_\ell}\RA^{(1,2,3)}$ consisting of bi-local operators in the $\cX$ representation.

\paragraph{$\cX$ representation}

First, consider the case $\ell=0$, given by the structure \eqref{TTO1P}.  
A significant benefit of the $\cX$ representation is that, after using the shadow relation \eqref{PXshadow}, the combination of normal-ordered operators in \eqref{eq:B1l0EP} turns into a purely algebraic factor:
\begin{align}
    \ :\!(\EE{12}\XX{12}-\DD{12}\DD{21})^2\!:\ =\hat H_{12}^2\,,
\end{align}
which follows from the identities \eqref{DDid3} and \eqref{DDid4}. 
The resulting expression for $\LA TT\O\RA$ in terms of the normalized $\cH_{12}$ is
\begin{equation}
\LA\wh{TT\O}\RA\, \propto\, \cH_{12}^2    - 2  \cH_{12}\EE{12}  + \frac{d-2}{d-1} \EE{12}^2\, .
\label{TTO}
\end{equation}
We see that this is almost given by the double copy of $\LA \wh{JJ\O}\RA$, except the slightly funny coefficient for the last term. 
The flat-space limit is thus only partially manifest in this representation. 
Despite this, evaluating this expression is extremely simple, and this approach will similarly prove beneficial in the subsequent examples.

\definecolor{blue3}{RGB}{31, 119, 180}
\definecolor{red3}{RGB}{214, 39, 40}
\definecolor{orange3}{RGB}{255, 127, 14}
\definecolor{green3}{RGB}{44, 160, 44}
\definecolor{Purple}{RGB}{178, 102, 255}

\newcommand{\teal}[1]{\textcolor{teal}{#1}} 
\newcommand{\magenta}[1]{\textcolor{magenta}{#1}}
\newcommand{\blue}[1]{\textcolor{blue3}{#1}}
\newcommand{\green}[1]{\textcolor{green3}{#1}}
\newcommand{\red}[1]{\textcolor{red3}{#1}}
\newcommand{\purple}[1]{\textcolor{Purple}{#1}}
\newcommand{\orange}[1]{\textcolor{orange3}{#1}}
\newcommand{\brown}[1]{\textcolor{brown}{#1}}

For $\ell\ge 4$, there are three independent three-point structures for $\LA TT\O_\ell\RA$ in $d>3$ (and two in $d=3$).
To generate these structures, we need to employ a larger set of differential operators than in \eqref{TTO}.
There are in total of ten independent differential building blocks that are symmetric between 1 and~2, and a convenient basis is given by
\beq
\begin{aligned}
   \mathbb{O}_1  & = \red{\cH_{12}^2} \,,  &&\mathbb{O}_6  =  \green{\cH_{12}\cE_{13}\cE_{23}}\,,\\
   \mathbb{O}_2  & = \red{\cH_{12}\cE_{12}} \,,  \qquad &&\mathbb{O}_7 = \green{\cE_{12}\cE_{13}\cE_{23}}\,,\\
   \mathbb{O}_3  & = \red{\cE_{12}^2} \,,  \qquad &&\mathbb{O}_8 = \green{\cE_{13}^2\cD_{21}^2+\cE_{23}^2\cD_{12}^2}\,,\\
   \mathbb{O}_4  & = \orange{\cH_{12}(\cE_{13}\cD_{21}+\cE_{23}\cD_{12})} \,,  \qquad &&\mathbb{O}_9 =\blue{\cE_{13}^2\cE_{23}\cD_{21}+\cE_{23}^2\cE_{13}\cD_{12}}\,,\\
   \mathbb{O}_5  & =
   \orange{\cE_{12}(\cE_{13}\cD_{21}+\cE_{23}\cD_{12})}\,,  \qquad &&\hskip -4pt \mathbb{O}_{10} = \purple{\cE_{13}^2\cE_{23}^2}\,.
\end{aligned}\label{Obasis2}
\eeq
These differential operators have been organized according to increasing powers of the spin-raising operator $\cE_3$. 
When $\mathbb{O}_I$ containing different numbers of $\cE_3$ combine, additional factors of $\cV_3$ is inserted in the seed.

Let us consider the general spin-$\ell$ expression for the first structure $\LA TT\O_\ell\RA^{(1)}$, which exists for all $\ell \ge 0$ and reduces to $\LA TT\cO \RA$ when $\ell=0$. 
In terms of the operators $\mathbb{O}_I$, this takes the form
\begin{align}
    \LA \wh{TT\O_\ell}\RA^{(1)}= \Big[\mathbb{O}_1c_1+\mathbb{O}_2c_2+\mathbb{O}_3c_3 &+\ell (\mathbb{O}_4c_4+\mathbb{O}_5c_5)\cV_3^{-1}\nn[-3pt]
    & +\ell(\ell-1)(\mathbb{O}_6c_6+\mathbb{O}_7c_7+\mathbb{O}_8c_8)\cV_3^{-2}\nn
    & +\ell(\ell-1)(\ell-2) \mathbb{O}_9c_9\cV_3^{-3}\nn
    &+\ell(\ell-1)(\ell-2)(\ell-3) \mathbb{O}_{10}c_{10}\cV_3^{-4}\Big]\cV_3^\ell\,,
    \label{eq:M1genldown}
\end{align}
where
\beq
\begin{aligned}
    c_1&=d^2(d-1)^2\,, \quad && c_6 =4d^2(d-1)^2 \,,\\
    c_2&=-2d(d-1)^2(d+2\ell)\,, && c_7 =-2d^2((d-2)(d-4+2\ell)+\cX_{12})\,,\\
    c_3&= d(d-1)(d+2\ell)(d-2+2\ell)\,,\quad && c_8 = d(d-2)\,,\\
    c_4&=-2d(d-1)^2\,, && c_9 = -2d^2(d-2)\,,\\
    c_5&=2d(d-1)(d-2+2\ell)\,, && c_{10} = d^2(d-2)^2\,.
\end{aligned}
\eeq
We have included $\cX_{12}$ in $c_7$ for convenience, which gives a factor of $\frac{1}{2}(d-2+\Delta+\ell)(2d-2-\Delta+\ell)$ when acting on the seed function $\cV_3^{\ell-2}\LA\phi_d\phi_d\phi_{\Delta+\ell-2}\RA$. 
We see that all coefficients presented this way are then manifestly independent of the dimension of $\O_\ell$, and that only the first three terms survive when $\ell=0$, reproducing \eqref{TTO}. 
Despite the seemingly more complex coefficients, evaluating this expression is highly efficient compared to \eqref{TTO1P}.

In $d>3$, the second structure $\LA TT\cO_\ell\RA^{(2)}$ is also non-vanishing for $\ell\ge 2$. It can be distinguished from the first structure $\LA TT\O_\ell\RA^{(1)}$ by the non-presence of the term $\mathbb{O}_1=\cH_{12}^2$. 
We find
\begin{align}
 \LA\wh{TT\O_\ell}\RA^{(2)}  = \Big[ \mathbb{O}_2e_2+\mathbb{O}_3e_3  &+ (\mathbb{O}_4e_4 +\mathbb{O}_5 e_5)\cV_3^{-1} +(\mathbb{O}_6e_6+\mathbb{O}_7e_7+\mathbb{O}_8e_8)\cV_3^{-2}     \nn
&\qquad  + (\ell-2)\mathbb{O}_9 e_9 \cV_3^{-3} + (\ell-2)(\ell-3) \mathbb{O}_{10}    e_{10}\cV_3^{-4} \Big] \cV_3^{\ell} \label{eq:M2genldown} \ ,
\end{align}
where second line only contributes when $\ell>2$. The coefficients are given by
\beq
\begin{aligned}
    e_2&=(d-1)^2\,, \quad && e_7 =  -d^2((d-3)(d-4+2\ell)-(\ell-2)(\ell-3))\\
    e_3&= -d(d-2+2\ell)\,, && \qquad + d(d-8+4\ell) \cX_{12} \,,\\
    e_4&= (d-1)^2\,,\quad && e_8 = -d(\ell-1) \,,\\
    e_5&= -d(d-4+3\ell)\,, && e_9 = -d^2(d-\ell) + d\cX_{12}\,,\\
    e_6&= d(d-1)^2 d(d+4-2\ell-\cX_{12})\,, && e_{10} = \tfrac{1}{2}d^2 ((d-2)^2 - \cX_{12}) \,.
\end{aligned}
\eeq
These coefficients again become independent of the dimension of $\O_\ell$, after absorbing some of them to $\cX_{12}$.

Lastly, there is a third structure $\LA TT\O_\ell\RA^{(3)}$ that exists for $\ell\ge 4$. Instead of using $\mathbb{O}_I$, we find it more convenient to express this structure as
\begin{align}
    \LA\wh{TT\O_\ell}\RA^{(3)} &= \ :\! (\cU^2 -2d^2 \cU\EE{13}\EE{23} + d^2(d-1)(d-2)(\EE{13}\EE{23})^2)\!: \cV_3^{\ell-4}\,,\label{eq:M3genldown}\\
    \cU &\equiv \EE{13} \EE{23} \XX{12} + \EE{12}\DD{31}\DD{32}   - \EE{23}\DD{12}\DD{31} - \EE{13} \DD{21} \DD{32} \, .
\end{align}
The extra $d$-dependent terms compared to \eqref{TTO3P} can be understood as arising from the shadow relation, whose coefficients are proportional to $\Delta_T=d$ instead of $\tDelta_T=0$ in \eqref{TTO3P}. 
While \eqref{eq:M3genldown} is not expressed purely in terms of bi-local operators, evaluating this is still very efficient. 
It is also straightforward to write this in terms of $\mathbb{O}_I$ with, this time, $\Delta$-dependent coefficients.

It can be straightforwardly verified that all three structures $\LA TT\O_\ell\RA^{(1, 2, 3)}$ are conserved for their valid ranges of $\ell$. 
Although these $\cX$ representations may appear somewhat intricate, these structures need to be evaluated only once, which can then be expressed in the algebraic basis consisting of $H_{ij}$ and $V_{i,jk}$.
The \textsc{Mathematica} notebook containing the explicit expressions of $\LA TT\O_\ell\RA^{(1,2,3)}$ can be found at~\href{https://github.com/haydenhylee/ampbasis}{\faGithub}.

\subsection{$\langle J \mathcal{O}_\ell \mathcal{O}_\ell \rangle$ and $\langle T \mathcal{O}_\ell \mathcal{O}_\ell \rangle$}\label{sec:JOOTOO}


Next, we examine three-point structures involving one conserved tensor and two non-conserved spinning tensors. 

\subsubsection{$\langle J \mathcal{O}_\ell \mathcal{O}_\ell \rangle$}

For the scattering of one photon and two charged massive spinning particles, there are a total of $(\ell+1) + \ell =2\ell+1$ independent amplitudes that are gauge invariant in $\eps_1$ and antisymmetric under the exchange $2\leftrightarrow 3$. 
The two groups of on-shell amplitudes are
\begin{align}
    \cA_{1\ell\ell}^{(1,m)} &= \ee{23}^{\ell-m}(\dd{23}\dd{32})^m\dd{12} &&(0\le m\le \ell)\,, \label{eq:JOOA1genl}\\
     \cA_{1\ell\ell}^{(2,m)} &= \ee{23}^{\ell-m-1}(\dd{23}\dd{32})^m(\ee{12}\dd{31}-\ee{13}\dd{21}) &&(0\le m\le \ell-1)\,. \label{eq:JOOA2genl}
\end{align}
The corresponding Lagrangian structures are shown in Table~\ref{tab:JOO}.

\begin{table}
\begin {center}
\begin {tblr} { | l | l | l |  } \hline
Dim. & Lagrangian & Spin  \\ \hline
$D \geq 4$ & $  A^\mu\nabla_{\rho_1 \cdots \rho_m} S^{\nu_1 \cdots \nu_\ell}\nabla_{\mu\nu_1 \cdots \nu_m} S^{\star\rho_1 \cdots \rho_m}{}_{\nu_{m+1} \cdots \nu_\ell}-\text{c.c.}$  & $0\leq m \leq \ell$ \\ \hline[dotted]
$D \geq 4$  & \begin {tabular}{@{}l@{}} $ A_{\nu_1} \nabla_{\rho_1 \cdots \rho_{m+1}} S^{\nu_1 \cdots \nu_\ell}\nabla_{\nu_2 \cdots \nu_{m+1}}S^{\star\rho_1 \cdots \rho_{m+1}}{}_{\nu_{m+2} \cdots \nu_\ell}$ \\ $-  A_{\rho_1}\nabla_{\rho_2 \cdots \rho_{m+1}} S^{\nu_1 \cdots \nu_\ell} \nabla_{\nu_1 \cdots \nu_{m+1}} S^{\star\rho_1 \cdots \rho_{m+1}}{}_{\nu_{m+2} \cdots \nu_\ell}-\text{c.c.}$ \end {tabular} & $0 \leq m \leq \ell-1$ \\ \hline
\end {tblr}
\caption{Cubic interactions for one photon and two massive spinning particles. There are in total of $2\ell+1$ independent parity-even amplitudes in $D\ge 4$, classified into two groups.}
\label{tab:JOO}
\end {center}
\end{table}

An important case is the cubic interaction from the minimally-coupled kinetic term, $A^\mu S^{\nu_1 \cdots \nu_\ell} \nabla_\mu S^\star_{\nu_1 \cdots \nu_\ell}-\text{c.c.}$,
where $A^\mu$ is a spin-1 gauge field. 
 This gives the amplitude $\cA_{1\ell\ell}^{(1,0)} = \epsilon_{23}^\ell d_{12}$. 
The corresponding conformal three-point structure is easy to write down:
\begin{align}
    \LA \wh{J\O_\ell\O_\ell}\RA^{(1,0)} = 
    \EE{23}^\ell \cV_1\,. \label{JOOmin}
\end{align}
Since $\text{div}_1\cV_1=0$, this structure is manifestly conserved at position 1 at non-coincident points, while the Ward identity relates the coefficient of this structure to the two-point function of $\O_\ell$. 
As an example, evaluating the result for $\ell=1$ and writing it in the algebraic basis gives
\begin{align}
    \LA J\O_1\O_1\RA^{(1,0)} \propto \frac{H_{12}V_3+H_{13}V_2+ (d-2)V_1V_2V_3-(1+\frac{2\Delta(\Delta-1)}{d-2})H_{23}V_1}{X_{12}^{d/2}X_{13}^{d/2}X_{23}^{\Delta+1-d/2}}\,.
\end{align}
Because $\cV_1$ depends on both $X_2$ and $X_3$, we see that the result after acting with $\cE_{23}^\ell$ is not simply proportional to $V_1$, obscuring the flat-space limit. 
This makes it difficult to distinguish this structure from the ones associated to higher-derivative interactions, suppressed by $\Delta_{\rm gap}$.
In contrast, the differential representation \eqref{JOOmin} clearly displays the flat-space structure.

In the general case, we can again uplift the amplitudes \eqref{eq:JOOA1genl} and \eqref{eq:JOOA2genl} directly to write down the corresponding conformal structures in the $\cP$ representation. However, evaluating them turns out to be highly nontrivial for general $\ell$ and $m$, and therefore this time we only explicitly present the expressions in the $\cX$ representation, given by
\begin{align}
    \LA \wh{J\O_\ell\O_\ell}\RA^{(1,m)} &= \EE{23}^{\ell-m}(\DD{23}\DD{32})^m\cV_1\,, \label{eq:JOOA1down}\\
    \LA \wh{J\O_\ell\O_\ell}\RA^{(2,m)} &= \EE{23}^{\ell-m-1}(\DD{23}\DD{32})^m (\EE{12}\DD{31}-\EE{13}\DD{21})\,.\label{eq:JOOA2down}
\end{align}
It should be noted that this is not a unique basis choice that one can use. 
Instead of the insertion $(\DD{23}\DD{32})^m= \DD{23}^m\DD{32}^m$, the one that may more naturally arise from the direct uplifting of the amplitude is
\begin{equation}
    :\!(\DD{23}\DD{32})^m\!: \,= \cE_2^{A_1}\cdots\cE_2^{A_m}\cE_3^{B_1}\cdots\cE_3^{B_m}\cX_{3A_1}\cdots \cX_{3A_m}\cX_{2B_1}\cdots \cX_{2B_m}\,.
\end{equation}
It is clear that this replacement does not affect the conservation at position 1. 
However, this proliferation of indices is difficult to handle for high $m$, and we have opted for the representation that is the simplest for evaluation.
This non-uniqueness of the basis choice follows from the non-commutative nature of covariant derivatives in AdS, and there are multiple ways of organizing the AdS interactions that involve shuffling around the lower-derivative terms.

\subsubsection{$\langle T \mathcal{O}_\ell \mathcal{O}_\ell\rangle $}

For the scattering of one graviton and two identical massive spinning particles, we have a total of $(\ell+1) + \ell + (\ell-1) = 3 \ell$ independent amplitudes that are gauge invariant in $\eps_1$ and symmetric under the exchange $2 \leftrightarrow 3$. In $D>4$, the three groups of amplitudes are
\begin{align}
    \cM_{2\ell\ell}^{(1,m)} &= (\dd{23}\dd{32})^m\ee{23}^{\ell-m}\dd{12}^2&& (0\le m\le \ell)\,, \label{eq:TOOM1genl}\\
    \cM_{2\ell\ell}^{(2,m)} &=  (\dd{23}\dd{32})^m\ee{23}^{\ell-m-1}\dd{12}(\ee{12}\dd{31}-\ee{13}\dd{21})&& (0\le m\le \ell-1)\,, \label{eq:TOOM2genl}\\
    \cM_{2\ell\ell}^{(3,m)} &=  (\dd{23}\dd{32})^m\ee{23}^{\ell-m-2}(\ee{12}\dd{31}-\ee{13}\dd{21})^2 && (0\le m\le \ell-2)\,.\label{eq:TOOM3genl}
\end{align}
Only the first two groups of amplitudes exist in $D=4$ spacetime dimensions. The corresponding Lagrangian structures are shown in Table \ref{tab:TOO}.

\begin{table}
\begin {center}
\begin {tblr} { | l | l | l |  } \hline
Dim. & Lagrangian & Spin  \\ \hline
$D \geq 4$ & $  h^{\mu_1 \mu_2} \nabla_{\mu_1 \rho_1 \cdots \rho_m} S^{\nu_1 \cdots \nu_\ell} \nabla_{\mu_2\nu_1 \cdots \nu_m} S^{\rho_1 \cdots \rho_m}{}_{\nu_{m+1} \cdots \nu_\ell}$  & $ 0\leq m \leq \ell$ \\ \hline[dotted]
$D \geq 4$ & \begin {tabular}{@{}l@{}} $ h^{\mu_2}{}_{\nu_1} \nabla_{\rho_1 \cdots \rho_{m+1}} S^{\nu_1 \cdots \nu_\ell} \nabla_{\mu_2\nu_2 \cdots \nu_{m+1}} S^{\rho_1 \cdots \rho_{m+1}}{}_{\nu_{m+2} \cdots \nu_\ell}$ \\ $-  h^{\mu_2}{}_{\rho_1} \nabla_{\rho_2 \cdots \rho_{m+1}} S^{\nu_1 \cdots \nu_\ell} \nabla_{\mu_2\nu_1 \cdots \nu_{m+1}} S^{\rho_1 \cdots \rho_{m+1}}{}_{\nu_{m+2} \cdots \nu_\ell}$ \end {tabular} &  $0 \leq m \leq \ell-1$ \\ \hline[dotted]
$D \geq 5$ & \begin {tabular}{@{}l@{}} $ h_{\nu_1 \nu_2} \nabla_{\rho_1 \cdots \rho_{m+2}} S^{\nu_1 \cdots \nu_\ell} \nabla_{\nu_3 \cdots \nu_{m+2}}S^{\rho_1 \cdots \rho_{m+2}}{}_{\nu_{m+3} \cdots \nu_\ell}$ \\ $- 2  h_{\nu_1 \rho_1} \nabla_{\rho_2 \cdots \rho_{m+2}} S^{\nu_1 \cdots \nu_\ell} \nabla_{\nu_2 \cdots \nu_{m+2}}S^{\rho_1 \cdots \rho_{m+2}}{}_{\nu_{m+3} \cdots \nu_\ell}$ \\ $+ h_{\rho_1 \rho_2} \nabla_{\rho_3 \cdots \rho_{m+2}} S^{\nu_1 \cdots \nu_\ell}\nabla_{\nu_1 \cdots \nu_{m+2}} S^{\rho_1 \cdots \rho_{m+2}}{}_{\nu_{m+3} \cdots \nu_\ell}$  \end{tabular} & $0\leq m \leq \ell-2 $ \\ \hline
\end {tblr}
\caption{Cubic interactions for one graviton and two identical massive spinning particles. There are in total of $3\ell$ independent parity-even amplitudes in $D\ge 5$ ($2\ell+1$ in $D = 4$), classified into three (two in $D=4$) groups.}
\label{tab:TOO}
\end {center}
\end{table}

Before presenting the general formulas, let us first consider the minimal coupling of the kinetic term, $h^{\mu_1 \mu_2}\nabla_{\mu_1} S^{\nu_1 \cdots \nu_\ell} \nabla_{\mu_2} S_{\nu_1 \cdots \nu_\ell}$, where $h^{\mu_1 \mu_2}$ denotes the graviton. 
This gives the amplitude $\cM_{2\ell\ell}^{(1,0)} = \epsilon_{23}^\ell d_{12}^2$, with the corresponding uplifted expression $\EE{23}\cV_1^2$.
We have already described the relevant features of a similar example \eqref{JOOmin}, and here we also show the explicit result for $\ell=1$:
\begin{align}
  \LA T\O_1\O_1 \RA^{(1,0)}  &= \EE{23} \cV_1^2\langle \phi_{d+2} \phi_{\Delta} \phi_{\Delta} \rangle\\
&\propto \frac{H_{12}H_{13} -(d-2)(H_{13}V_2+H_{12}V_3)V_1+\frac{(d-2)^2}{2}V_2V_3V_1^2 + \frac{2\Delta(\Delta-1)+d-2}{2}H_{23}V_1^2}{X_{12}^{d/2+1}X_{13}^{d/2+1}X_{23}^{\Delta-d/2}}\,. \nonumber
\end {align}
We see that the flat-space structure is manifest in the differential representation but not so much in the algebraic basis. This structure stands out because it satisfies the Ward identity, and consequently its normalization is fixed by the two-point function of $\O_\ell$.

As before, we can write down 
the $\cP$ representation for general structures, but this approach is not ideal for practically computing the correlators with general $\ell$ and $m$. 
Instead, we simply show the expressions in the $\cX$ representation below:
  \begin{align}
     \LA\wh{T\O_\ell\O_\ell}\RA^{(1,m)} &= \EE{23}^{\ell-m}(\DD{23}\DD{32})^m\cV_1^2\,\label{eq:TOOM1down},\\
     \LA\wh{T\O_\ell\O_\ell}\RA^{(2,m)} &= \EE{23}^{\ell-m-1}(\DD{23}\DD{32})^m\big((\EE{12}\DD{31}-\EE{13}\DD{21})\cV_1-d \EE{12}\EE{13}\big)\,\label{eq:TOOM2down}, \\
     \LA\wh{T\O_\ell\O_\ell}\RA^{(3,m)} &= \EE{23}^{\ell-m-2} (\DD{23}\DD{32})^m \big(\!:\!(\EE{12}\DD{31}-\EE{13}\DD{21})^2\!:-\,2d\EE{12}\EE{13}\EE{23}\big)\, \label{eq:TOOM3down}.
 \end{align}
Notice the presence of the extra terms that are proportional to $d$. As we have seen in \eqref{eq:M3genldown}, these $d$-dependent coefficients can be understood as arising from the shadow relation, and are required to satisfy the conservation constraint.

\section{Spinning Conformal Blocks}\label{sec:spinningblocks}

Having constructed a basis for three-point tensor structures, we now outline a method to construct the corresponding basis of conformal blocks. 
An efficient way to compute spinning conformal blocks is using the method in \cite{Costa:2011dw}, which involves applying differential operators on scalar seed blocks. 
We review this method and give the differential representations for the conformal block decomposition of $\LA JJJJ\RA$ and $\LA TTTT\RA$ in terms of the amplitude basis of three-point structures we presented in the previous section.

\subsection{Differential Representation and Seed Blocks}\label{sec:blocks}

To begin with, consider the four-point function of a scalar $\phi$. 
By summing over all primary operators that contribute to the OPE in the 12-34 channel, the four-point function can be expressed as 
\begin{align}
  \LA \phi(X_1)\phi(X_2)\phi(X_3)\phi(X_4)\RA
  &= \mathbb{K}_4(X_i)\sum_{\O} \lambda_{\phi\phi\O}^2 G_\O(u,v)\,,
\end{align}
where $\lambda_{\phi\phi\O}$ denotes the OPE coefficient, the conformal block $G_\O$ is a dimensionless function of the conformally-invariant cross-ratios
\begin{align}
    u=\frac{X_{12}X_{34}}{X_{13}X_{24}}\,,\quad v=\frac{X_{14}X_{23}}{X_{13}X_{24}}\,,
\end{align}
and the overall kinematical factor $\mathbb{K}_4$ is given by
\begin{align}
    \mathbb{K}_4(X_i)& =\left(\frac{X_{24}}{X_{14}}\right)^{\frac{\Delta_1-\Delta_2}{2}}\left(\frac{X_{14}}{X_{13}}\right)^{\frac{\Delta_3-\Delta_4}{2}} \frac{1}{(X_{12})^{\frac{\Delta_1+\Delta_2}{2}}(X_{34})^{\frac{\Delta_3+\Delta_4}{2}}}\,.
\end{align}
The conformal block satisfies the conformal Casimir equation 
\begin{equation}
	(\cC_2-C_{\Delta,\ell})G_\O=0\,,
\end{equation} 
where $\cC_2=(\cL_1+\cL_2)^2$ is the quadratic Casimir of the conformal group and $C_{\Delta,\ell}=\Delta(\Delta-d)+\ell(\ell+d-2)$ is the Casimir eigenvalue for the exchanged operator $\O$.
The scalar conformal block is well known, which can be computed by solving the Casimir equation and imposing the correct behavior in the OPE limit~\cite{Dolan:2003hv}.

Similar to the scalar case, a spinning four-point function can be expanded in conformal blocks together with conformally-invariant tensor structures. 
Denoting the external operators as $\O_i\equiv\O_{\ell_i}(Z_i,X_i)$, the expansion of their four-point function in the 12-34 channel gives
\begin{align}
	\LA \O_1\O_2\O_3\O_4\RA&=\mathbb{K}_4(X_i)\sum_\O\sum_{a,b} \lambda_{\O_1\O_2\O}^{(a)}\lambda_{\O_3\O_4\O}^{(b)}G_{\O}^{(a,b)}(Z_i,X_i) \nn
 &=\mathbb{K}_4(X_i)\sum_\O\sum_{a,b}\lambda_{\O_1\O_2\O}^{(a)}\lambda_{\O_3\O_4\O}^{(b)}\sum_I \mathbb{T}_{\O_1\cdots\O_4}^I(Z_i,X_i) G_{\O}^{I,(a,b)}(u,v)\,,\label{Texp}
\end{align}
where $\mathbb{T}_{\O_1\cdots\O_4}^I$ denotes conformally-invariant four-point tensor structures indexed by $I$, and dimensionless functions $G_{\O}^{I,(a,b)}$ are referred to as the spinning conformal blocks.
This expansion involves a double sum over $a,b$ that label the independent three-point structures. 
An important point is that, unlike the scalar version, spinning conformal blocks depend on the choice of a basis for three-point tensor structures. 
A well-chosen basis can therefore facilitate the computation and also clarify the interpretation of the OPE data.

An efficient way to compute these spinning blocks is by acting with differential operators  on scalar conformal blocks as~\cite{Costa:2011dw}
\begin{align}
    \mathbb{K}_4(X_i) G_{\O}^{(a,b)}(Z_i,X_i) &= \cD^{(a)}_{\O_1\O_2}\cD^{(b)}_{\O_3\O_4}  \mathbb{K}_4(X_i) G_{\O}(X_i)\,,\label{KGspin}
\end{align}
where $\cD_{\cO_i \cO_j}^{(a)}$ are some differential operators and the scalar conformal block $G_{\O}(X_i)=G_{\O}(u,v)$ here is referred to as a seed block.
One can generate the desired tensor structures with a linear combination of products of differential operators, i.e.,
\begin{align}\label{DO1O2}
    \cD^{(a)}_{\O_1\O_2} \supset \cH_{12}^{n_{12}}\FF{12}^{m_1}\FF{21}^{m_2}\DD{12}^{m_1'}\DD{21}^{m_2'}\Sigma^{m_1+m_2'+n_{12},m_2+m_1'+n_{12}}\,,
\end{align}
where $\{n_{12}+m_1+m_1',n_{12}+m_2+m_2'\}=\{\ell_1,\ell_2\}$  and $\Sigma^{\delta_1,\delta_2}: \{\Delta_1,\Delta_2\}\to\{\Delta_1+\delta_1,\Delta_2+\delta_2\}$ is a reminder that the scaling dimensions of the seed block should be increased accordingly to account for the dimension shifts from the weight-shifting operators.
For identical external operators, $\cD_{\O_3\O_4}^{(a)}$ has the same form  as $\cD_{\O_1\O_2}^{(a)}$ under the permutation $1,2\leftrightarrow 3,4$. 
Following the practice in the previous section, we will henceforth suppress the dependence on $\Sigma^{\delta_1,\delta_2}$ in~$\cD_{\O_1\O_2}^{(a)}$.

The problem of computing spinning conformal blocks boils down to constructing conformally-invariant differential operators that satisfy
\begin{align}
    \LA  \O_1\O_2 \O_\ell(Z_3,X_3)\RA^{(a)} &= \cD_{\O_1\O_2}^{(a)}(Z_1,Z_2,X_1,X_2) \LA \phi(X_1)\phi(X_2) \O_\ell(Z_3,X_3)\RA\nn
    &= \cD_{\O_1\O_2}^{(a)}(Z_1,Z_2,X_1,X_2)\cV_3^\ell \LA \phi(X_1)\phi(X_2) \phi_{\Delta+\ell}(X_3)\RA\,,\label{Dseed}
\end{align}
where $\ell$ denotes the spin of the exchanged operator, and we have shown some arguments of the operators for clarification. 
A crucial property of $\cD_{\O_1\O_2}^{(a)}$ is that it only acts on external scalars, namely $\phi(X_1)$ and $\phi(X_2)$ in \eqref{Dseed}, so that it does not affect the conformal multiplet of the exchanged operator in the OPE.
These scalars in the seed blocks are not necessarily operators present in a CFT. Instead, they are considered for pure kinematical reasons, whose three-point functions are normalized as in \eqref{seed}.

In Section~\ref{sec:ampbasis}, we have defined a basis of three-point structures for $\LA JJ\O_\ell\RA$ and $\LA TT\O_\ell\RA$. 
Recall that some differential operators for these structures acted on $X_3$, which naturally followed from the uplifting of amplitudes.
To use the above method for computing spinning conformal blocks, we would like to translate the results of the previous section into differential operators that just act on $X_1$ and $X_2$. 
We carry out this exercise next.

\subsection{$\LA JJJJ\RA$ and $\LA TTTT\RA$}\label{sec:blocks}

We will now present two concrete examples, focusing on the conformal blocks for the four-point functions $\LA JJJJ\RA$ and $\LA TTTT\RA$ due to the exchange of totally symmetric tensors.

\subsubsection{$\LA JJJJ\RA$}

For the conformal blocks with external conserved currents, we would like to derive the differential operators $\cD_{J_1J_2}^{(a)}$ that generate the three-point structures $\LA JJ\O_\ell\RA^{(a)}$ that we have classified before.
As before, there are four independent tensor structures symmetric in 1 and 2 before imposing conservation.
A set of four independent differential operators of the form \eqref{DO1O2} is given by
\beq
\begin{aligned}
   \mathbb{D}_1   & =\red{\cH_{12}} \,,  &&\mathbb{D}_3  = \green{\cF_{12}\cD_{21}+\cF_{21}\cD_{12}}\,,\\
   \mathbb{D}_2  & =\red{\cF_{12}\cF_{21}} \,,  \qquad &&\mathbb{D}_4 =\blue{\cD_{12}\cD_{21}}\,.
\end{aligned}
\eeq
Our task is to recast the amplitude basis structures $\LA JJ\O_\ell\RA^{(1,2)}$ in terms of $\mathbb{D}_I$. 
When doing so, the simplicity observed in one representation may not necessarily be reflected in the other.
For the purpose of facilitating conformal bootstrap applications, we would like the resulting expressions in terms of the $\mathbb{D}_I$ operators to be simple.
This leads us to perform a simple basis rotation for the three-point structures $\LA JJ\O_\ell\RA^{(1,2)}$, and define the two independent differential operators for the conformal blocks as follows:
\beq
\begin{aligned}
    \cD_{J_1J_2}^{(1)} \cV_3^\ell & \equiv \LA \wh{JJ\O_\ell}\RA^{(1)}+\ell(\ell-1)\LA \wh{JJ\O_\ell}\RA^{(2)}\,,\\
    \cD_{J_1J_2}^{(2)}\cV_3^\ell & \equiv \LA \wh{JJ\O_\ell}\RA^{(2)}\,.
\end{aligned}
\eeq
Such a basis rotation for the first structure is in fact naturally motivated from the bulk perspective, where the exact three-point function computed in AdS is also a linear combination of $\LA JJ\O_\ell\RA^{(1)}$ and $\LA JJ\O_\ell\RA^{(2)}$, as we show in Appendix~\ref{app:bulk}.
The superscript label $a\in\{1,2\}$ matches the label we used for the amplitude basis structures, so that $\cD_{J_1J_2}^{(1)}$ and $\cD_{J_1J_2}^{(2)}$ exist for $\ell\ge 0$ and $\ell\ge 2$, respectively.
Explicitly, we have
\begin{align}
\cD_{J_1J_2}^{(1)} 
&\propto C_{\Delta,\ell} \mathbb{D}_1 + 2 \mathbb{D}_2=C_{\Delta,\ell}\cH_{12}+2\cF_{12}\cF_{21} \,,\label{JJODleft1}\\[3pt]
   \cD_{J_1J_2}^{(2)} &\propto  C_{\Delta,\ell} \left( \ell(\ell+1)\mathbb{D}_1+\mathbb{D}_2-\mathbb{D}_3+\frac{d-1}{d-2}\mathbb{D}_4\right)\nn
    &\quad -2(\ell-1)\Big((d-1+\ell)(d-2+\ell)\mathbb{D}_1+ (2d-4+\ell)\mathbb{D}_2 -(d-2+\ell)\mathbb{D}_3 \Big)\,.
\end{align}
The differential operators $\cD_{J_3J_4}^{(1,2)}$ are given by the permutation $1,2\leftrightarrow 3,4$. 
Interestingly, the first line is precisely the uplifted expression from the flat-space amplitude $ \cA_{11\ell}^{(1)} \propto  ( m^2\eps_{12}+2 d_{12}d_{21}  )d_{31}^\ell$ in terms of the bi-local operators in the adjoint representation, with the Casimir eigenvalue playing the role of the mass of a particle. 
In particular, unlike in \eqref{eq:A1genldown}, no extra terms need to be added for higher $\ell$. 
The appearance of the Casimir eigenvalue was also noticed as a curious observation in~\cite{Costa:2011dw}; here, we have given a physical interpretation for its appearance.

The differential operators we have presented are defined in terms of derivatives with respect to embedding-space coordinates $X_i$, which act on the combination of the kinematic factor and the scalar block, shown in \eqref{KGspin}. 
One can then evaluate the derivatives and express the resulting tensor structure $\mathbb{T}_{JJJJ}^I$ as a polynomial of the standard $H_{ij}$ and $V_{i,jk}$, and express the derivatives acting on the spinning conformal blocks in terms of the cross-ratio derivatives $\partial_u$ and $\partial_v$. 
Such a coordinate change can be straightforwardly worked out, see~\cite{Costa:2011mg,Dymarsky:2017xzb}.

\subsubsection{$\LA TTTT\RA$}

For $\LA TT\O_\ell\RA$ with $\ell\ge 4$, recall that there are 10 independent tensor structures symmetric in 1 and 2 before imposing conservation. 
These can be captured by the following set of $\mathbb{D}_I$ operators that act only on positions 1 and 2:
\beq
\begin{aligned}
   \mathbb{D}_1  & = \red{\cH_{12}^2} \,,  &&\mathbb{D}_6  =  \green{\cH_{12}\cD_{12}\cD_{21}}\,,\\
   \mathbb{D}_2  & = \red{\cH_{12}\cF_{12}\cF_{21}} \,,  \qquad &&\mathbb{D}_7 = \green{\cF_{12}\cF_{21}\cD_{12}\cD_{21}}\,,\\
   \mathbb{D}_3  & = \red{\cF_{12}^2\cF_{21}^2} \,,  \qquad &&\mathbb{D}_8 = \green{\cF_{12}^2\cD_{21}^2+\cF_{21}^2\cD_{12}^2}\,,\\
   \mathbb{D}_4  & = \orange{\cH_{12}(\cF_{12}\cD_{21}+\cF_{21}\cD_{12})} \,,  \qquad &&\mathbb{D}_9 =\blue{\cF_{12}\cD_{21}^2\cD_{12}+\cF_{21}\cD_{12}^2\cD_{21}}\,,\\
   \mathbb{D}_5  & = \orange{\cF_{12}^2\cF_{21}\cD_{21}+\cF_{21}^2\cF_{12}\cD_{12}}\,,  \qquad &&\hskip -4pt \mathbb{D}_{10} =\purple{\cD_{12}^2\cD_{21}^2}\,.
\end{aligned}
\eeq
These operators are organized in descending powers of $\cH$ and $\cF$, and are defined analogous to the $\mathbb{O}_I$ operators in~\eqref{Obasis2}. 
To avoid clutter, we use the same notation $\mathbb{D}_I$ as in the $\LA JJ\O_\ell\RA$ case to label the differential operators with the understanding that these are now associated with $\LA TT\O_\ell\RA$. 
Imposing conservation of $\LA TT \O_\ell \RA$ reduces the number of allowed combinations of $\mathbb{D}_I$ down to three.

As in the spin-1 case, it turns out that a certain basis rotation of $\LA TT\O_\ell\RA^{(a)}$ is useful for simplifying the resulting expressions in terms of the $\mathbb{D}_I$ operators.
For the first structure, we find that the following combination accomplishes this:
\begin{align}
    \cD_{T_1T_2}^{(1)} \cV_3^\ell  &\equiv  \wh\cT_1-\wh\cT_2 +\frac{(d-1)C_{\Delta,\ell}}{4(d-1+\ell)(d-2+\ell)}\left(\wh\cT_1-\wh\cT_2-\frac{d-2}{d-1}\wh\cT_3\right),
\end{align}
where
\begin{align}
    \wh\cT_1  &\equiv \LA\wh{TT\O_\ell}\RA^{(1)}\,,\\ \wh\cT_2  &\equiv \ell(\ell-1)\LA\wh{TT\O_\ell}\RA^{(2)}\,,\\ \wh\cT_3  &\equiv \ell(\ell-1)(\ell-2)(\ell-3)\LA\wh{TT\O_\ell}\RA^{(3)}\,.
\end{align}
This is so constructed that it becomes proportional to $\LA TT\O_\ell\RA^{(1)}$ for $\ell=0$, while keeping the coefficients of $\mathbb{D}_I$ relatively simple, with four of them set to zero.
Its explicit expression in terms of the $\mathbb{D}$ operators is given by
\begin{align}
    \cD_{T_1T_2}^{(1)}&\propto a_1\mathbb{D}_1+a_2\mathbb{D}_2+a_3\mathbb{D}_3+a_4\mathbb{D}_4+a_5\mathbb{D}_5+a_8\mathbb{D}_8\,,
\end{align}
where
\beq
\begin{aligned}
    a_1 &= \tfrac{1}{4}d(y_1+ y_2(C_{\Delta,\ell}-8))\,,\quad &&a_4 =  4d(2(\ell-1)(d-2+\ell)-C_{\Delta,\ell})\,,\\
    a_2 &=d(y_1+y_2+32(d-1))\,,\quad &&a_5 = -\tfrac{1}{2}a_4 \,,\\
    a_3 &= d((d-2)C_{\Delta,\ell}-4d^2)\,,\quad &&a_8 = 4(d+1)(2d-C_{\Delta,\ell})\,,\\
\end{aligned}
\eeq
with
\beq
\begin{aligned}
    y_1 &\equiv (d-1)(2d+4-C_{\Delta,\ell})(4d-8-C_{\Delta,\ell})\,,\\
    y_2 &\equiv 4\ell(d-2+\ell)((\ell-1)(d-1+\ell)-C_{\Delta,\ell})\,.
\end{aligned}
\eeq
For $\ell=0$, some $\mathbb{D}_i$ operators become degenerate and only three of them are linearly independent. 
We can make this fact manifest by rewriting the result for $\ell=0$ in terms of just $\mathbb{D}_{1,2,3}$ as 
\begin{align}
    \cD_{T_1T_2}^{(1)}\big|_{\ell=0} \,\propto\, C_{\Delta,0}^2\cH_{12}^2 + 4C_{\Delta,0}\cH_{12}\cF_{12}\cF_{21} + 4\,\frac{d-2}{d-1}\frac{\tilde\Delta\Delta}{(\tilde\Delta+2)(\Delta+2)}\cF_{12}^2\cF_{21}^2\, .\label{TTOL1HF2}
\end{align}
Notice that, similar to what we observed for $\LA TT\O\RA$ in \eqref{TTO}, this expression is almost given by the double copy of \eqref{JJODleft1}, except for one of the terms.
This is simply due to our desire to adopt the computationally efficient bi-local operators to express the result, while the double-copy structure can be made manifest in the $\cP$ representation, shown in \eqref{TTO1P}.

The remaining two independent differential operators $\cD_{T_1 T_2}^{(2, 3)}$ can be chosen as
\begin{align}
    \cD_{T_1T_2}^{(2)}\cV_3^\ell &\equiv \wh \cT_2+\left(\frac{1}{d-1}-1\right)\wh\cT_3\,,\\[-2pt]
    \cD_{T_1T_2}^{(3)}\cV_3^\ell &\equiv \wh 
    \cT_3\,.
\end{align}
For $\cD_{T_1T_2}^{(2)}$, which exists for $\ell\ge 2$ and $d\ge 4$, this combination of $\wh\cT_2$ and $\wh\cT_3$ simplifies some coefficients of $\mathbb{D}_I$, setting two of them to zero. 
The third structure $\cD_{T_1T_2}^{(3)}$ is chosen to be the same as $\wh\cT_3$ and it is thus non-vanishing only for $\ell\ge 4$. 
The explicit expressions for $\cD_{T_1T_2}^{(2, 3)}$ are slightly lengthier than that of $\cD_{T_1T_2}^{(1)}$, and we instead present them in the accompanying \textsc{Mathematica} notebook~\href{https://github.com/haydenhylee/ampbasis}{\faGithub}.

\section{Conclusions}\label{sec:con}

Stress tensors are universal in CFTs. By the AdS/CFT correspondence, they are dual to gravitons in the bulk AdS. 
Given this nature, it is highly important to gain a better understanding of the properties of stress tensor correlation functions in CFTs.
Stress tensors also hold a promising potential in the numerical bootstrap program, offering the possibility to enhance the constraints on CFTs beyond those obtained from scalar correlators.

In this paper, we have outlined a method to construct conformally-invariant three-point tensor structures in CFTs.
A key element of our approach is the use of weight-shifting operators~\cite{Karateev:2017jgd}. 
With appropriate normalization and ordering, these operators can be identified as the (A)dS analogs of the on-shell building blocks for scattering amplitudes, a relationship that can be made precise through the flat-space limit~\cite{Lee:2022fgr, Li:2022tby}. 
This understanding enables us to straightforwardly derive differential representations for tensor structures in CFT by uplifting flat-space amplitudes to AdS, involving simple substitutions of the building blocks.

Using this method, we have derived the differential representations for three-point structures involving conserved currents or stress tensors and non-conserved, totally symmetric tensors in $d\ge 3$. 
By construction, these differential representations precisely align with the established classification of scattering amplitudes involving photons/gravitons and massive spinning particles~\cite{Chakraborty:2020rxf}.
Such constructed amplitude basis of tensor structures then has a clear physical interpretation in holographic CFTs as corresponding to AdS correlators in one higher dimension.

While the uplifting procedure is straightforward and yields structures that satisfy the conservation constraints, the resulting differential representations, involving products of uncontracted weight-shifting operators, are often not optimal for explicit computations.
To address this, we have also presented computationally more efficient representations in terms of bi-local operators.
Since these operators can be easily Fourier transformed, the differential representations presented in this paper may also significantly streamline momentum-space calculations of three-point functions of conserved and non-conserved tensors~\cite{Gillioz:2018kwh,Isono:2019ihz,Marotta:2022jrp}. 
Our results for the three-point structures are summarized in Table~\ref{tab:summary}.

\begin{table}
\begin {center}
\begin{tblr} { | l | l | l | l | l | l |l |  } \hline
Dim. & Correlator &  Ref. & Amplitude & Ref. &Lagrangian & Ref. \\ \hline
 $d\ge 3$ & $\LA JJ\O_\ell\RA^{(1)}$ & \eqref{eq:A1genldown} &  $\cA_{11\ell}^{(1)}$ & \eqref{eq:JJOA1genl} &  $\nabla^\ell F F S$  & Tab.~\ref{tab:JJO} \\
 $d\ge 3$ & $\LA JJ\O_\ell\RA^{(2)}$ & \eqref{eq:A2genldown}&  $\cA_{11\ell}^{(2)}$ & \eqref{eq:JJOA2genl} & $\nabla^{\ell-2} F FS$ &  Tab.~\ref{tab:JJO}\\
 \hline[dotted]
 $d\ge 3$ & $\LA TT\O_\ell\RA^{(1)}$ &  \eqref{eq:M1genldown}& $\cM_{22\ell}^{(1)}$  & \eqref{eq:TTOM1genl} & $\nabla^{\ell} RR S$& Tab.~\ref{tab:TTO} \\
 $d\ge 4$ & $\LA TT\O_\ell\RA^{(2)}$ &  \eqref{eq:M2genldown}& $\cM_{22\ell}^{(2)}$ & \eqref{eq:TTOM2genl} &$\nabla^{\ell-2} RRS$ & Tab.~\ref{tab:TTO} \\
 $d\ge 3$ & $\LA TT\O_\ell\RA^{(3)}$ &  \eqref{eq:M3genldown}& $\cM_{22\ell}^{(3)}$ & \eqref{eq:TTOM3genl} & $\nabla^{\ell-4} RRS$& Tab.~\ref{tab:TTO}\\
 \hline[dotted]
 $d\ge 3$ & $\LA J\O_\ell \O_\ell\RA^{(1,m)}$ &  \eqref{eq:JOOA1down} & $\cA_{1\ell\ell}^{(1,m)}$ & \eqref{eq:JOOA1genl}& $ (A\nabla)\nabla^m S \nabla^{m} S^\star$  &  Tab.~\ref{tab:JOO}\\
 $d\ge 3$ & $\LA J\O_\ell \O_\ell\RA^{(2,m)}$ & \eqref{eq:JOOA2down} & $\cA_{1\ell\ell}^{(2,m)}$& \eqref{eq:JOOA2genl} & $ A\nabla^m S \nabla^{m+1}S^\star$  & Tab.~\ref{tab:JOO} \\
 \hline[dotted]
 $d\ge 3$ & $\LA T\O_\ell \O_\ell\RA^{(1,m)}$ & \eqref{eq:TOOM1down}  & $\cM_{2\ell\ell}^{(1,m)}$ & \eqref{eq:TOOM1genl} & $ (h\nabla^2) \nabla^m S \nabla^{m} S$ & Tab.~\ref{tab:TOO} \\
 $d\ge 3$ & $\LA T\O_\ell \O_\ell\RA^{(2,m)}$ & \eqref{eq:TOOM2down}  & $\cM_{2\ell\ell}^{(2,m)}$ & \eqref{eq:TOOM2genl} & $(h\nabla) \nabla^m S \nabla^{m+1} S$ & Tab.~\ref{tab:TOO} \\
 $d\ge 4$ & $\LA T\O_\ell \O_\ell\RA^{(3,m)}$ &  \eqref{eq:TOOM3down} & $\cM_{2\ell\ell}^{(3,m)}$ & \eqref{eq:TOOM3genl}  & $ h \nabla^{m} S \nabla^{m+2} S$ & 
 Tab.~\ref{tab:TOO} \\
 \hline
\end{tblr}
\caption{Summary of the three-point structures and schematic Lagrangians.}\label{tab:summary}
\end {center}
\end{table}

We have also examined the differential representations for spinning conformal blocks using the framework of~\cite{Costa:2011dw}, focusing on $\LA JJJJ\RA$ and $\LA TTTT\RA$ due to the exchange of totally symmetric tensors.
Interestingly, we have found that these spinning conformal blocks decompose somewhat nicely into the amplitude basis structures, suggesting that they could be useful for practical bootstrap applications beyond holographic CFTs. 
It would be fascinating to perform new bootstrap studies involving the stress tensor four-point function, building on the results of~\cite{Dymarsky:2017yzx} in $d=3$.
 
It is important to note that our analysis, although valid in general dimensions, was restricted to totally symmetric tensors and parity-even tensor structures.
This is clearly limited, as mixed-symmetric tensors can generally contribute to the OPE of stress tensors in ${d>3}$~\cite{Costa:2014rya,Costa:2016hju}.
Our results could be straightforwardly extended to incorporate these cases by uplifting the scattering amplitudes of two photons or gravitons and one mixed-symmetric tensor, which have also been classified in \cite{Chakraborty:2020rxf}.
An extension to supersymmetric and fermionic correlation functions would also be interesting.

In this work, we have refined the connection between weight-shifting operators in CFTs and flat-space scattering amplitudes that was more experimentally found in~\cite{Lee:2022fgr, Li:2022tby}.
We anticipate that our findings should have broad applications to holographic CFTs and the computation of (A)dS Witten diagrams. 
Specifically, the differential approach could significantly facilitate the calculation of graviton correlators and help to formulate a precise double copy in (A)dS.
In addition, it would be interesting to understand the relationship between the differential operators we have employed for conformal blocks and those applied to spinning partial waves of scattering amplitudes in~\cite{Buric:2023ykg}.

\paragraph{Acknowledgements} We thank Daniel Baumann, Clay C\'ordova, Liam Fitzpatrick, Austin Joyce, Matt Reece, and Sav Sethi for useful conversations.
HL is supported by the Kavli Institute for Cosmological Physics at the University of Chicago through an endowment from the Kavli Foundation and its founder Fred Kavli.

\appendix
\newpage
\section{Three-Point Functions in AdS}\label{app:bulk}

In the main text, we derived a basis for conformal three-point functions by uplifting the corresponding scattering amplitudes to AdS. 
A complementary (but more involved) approach is to derive them directly in the bulk AdS using perturbation theory.
In this appendix, we follow the latter approach and, through some basic examples, demonstrate that this gives the results presented in the main text.
We begin by describing the relevant details of spinning AdS propagators in~\S\ref{app:prop}, followed by a direct computation of $\LA JJ\mathcal{O}_\ell\RA$ for $\ell=0,2$ from the bulk perspective in~\S\ref{sec:JJObulk}.

\subsection{Spinning AdS Propagators}\label{app:prop}

In the main text, we have adopted the normalization convention for scalar seeds and weight-shifting operators that makes manifest their connection to scattering amplitudes in the flat-space limit within momentum space. 
This is sufficient for deriving differential representations for three-point structures in CFT, where overall numerical prefactors do not play an important role.

If one instead wants to perform an explicit calculation in AdS position space keeping precise coupling constants, then a different normalization choice might be more practical.
To accommodate this, we also present the normalization convention chosen in~\cite{Li:2022tby}, with a small modification. 
It is important to note, however, that all operator identities discussed in the main text and this appendix hold, regardless of the chosen normalization convention.

The standard normalization of the scalar bulk-to-boundary propagator in embedding space is given by 
\begin{align}
    \Pi_\Delta =  \frac{N_\Delta}{(-2X\cdot Y)^\Delta}\,,\quad N_\Delta = \frac{\Gamma(\Delta)}{2\pi^{\frac{d}{2}}\Gamma(\Delta-\frac{d}{2}+1)}\,.\label{scalarprop2}
\end{align}
A normalization choice for the weight-shifting operators consistent with \eqref{scalarprop2} is given by
\begin{align}
	\cX^A &= i(\tDelta-\Delta) \cD_{-0}^A \,,\label{cX2}\\
	\cE^A &= \frac{\Delta-\ell}{\Delta(\Delta-1)} \cD_{0+}^A \label{cE2}\,,\\[-3pt]
	\cP^A &= \frac{2i}{(\ell+1)(1-\Delta)(d-\Delta-2)(d-2\Delta-2)} \cD_{+0}^A \,.\label{cP2}
\end{align}
The bulk-to-boundary propagator for a  spin-$\ell$ field can then be obtained by
\begin{align}
\Pi_{\Delta}^{A_1\cdots A_\ell} & =  \cE^{A_1}\cdots \cE^{A_\ell}\Pi_\Delta\,.
\end{align}
Moreover, the operators $\cP$ and $\cX$ normalized as above also satisfy the same shadow relation given in~\eqref{PXshadow}.

In what follows, we provide some useful identities for the bulk-to-boundary propagators of scalars, spin-1, and spin-2 fields. 
These involve converting {\it bulk} covariant derivatives acting on a bulk-to-boundary propagator to {\it boundary} weight-shifting operators acting on scalar propagators, which allows us to pull the resulting derivatives with respect to external coordinates out of the bulk integral.

\paragraph{Scalar}
For bulk covariant derivatives acting on a scalar propagator, we have the following identities:
\begin{align}
	\nabla_A \Pi_\Delta &= i\cP_A \Pi_{\Delta-1} -\tDelta Y_A \Pi_\Delta\,,\nn
	&= i\cX_A \Pi_{\Delta+1} -\Delta Y_A\Pi_\Delta\,,\\
    	\nabla_A\nabla_B \Pi_\Delta &= -\cP_A\cP_B \Pi_{\Delta-2} + i(\tDelta-1)Y_{(A}\cP_{B)}\Pi_{\Delta-1} +\tDelta (\tDelta Y_AY_B-\eta_{AB})\Pi_\Delta\,, \nn
    	 &= -\cX_A\cX_B \Pi_{\Delta+2} + i(\Delta-1)Y_{(A}\cX_{B)}\Pi_{\Delta+1} +\Delta (\Delta Y_AY_B-\eta_{AB})\Pi_\Delta\,.
\end{align}
Note that the equivalent weight-down expressions are obtained by simply exchanging  $\cP^A\leftrightarrow \cX^A$, $\tDelta \leftrightarrow \Delta$, and accordingly shifting the weights of scalar seed propagators.

\paragraph{Spin one}

For a spin-1 field with generic weight $\Delta$, the repeated action of the covariant derivative on its propagator can be expressed as the combined action of spin-raising operator $\cE^A$ and the weight-raising $\cP^A$ on a collection of scalar propagators with suitably chosen weights. For instance,
\begin{align}
	(\Pi_{\Delta})_M &= \cE_M \Pi_{\Delta} \,,\label{eq:spin1prop}\\
	\nabla_A (\Pi_{\Delta})_M &= i\cE_M \cP_A \Pi_{\Delta-1} -(\tDelta Y_A \cE_M+Y_M \cE_A )\Pi_\Delta\,,\label{eq:covspin1}\\ 
 \nabla_A \nabla_B (\Pi_{\Delta})_M &=  - \mathcal{E}_{M} \mathcal{P}_{A} \mathcal{P}_{B} \Pi_{\Delta-2}- i (Y_M  \mathcal{E}_{(A} \mathcal{P}_{B)} + (\wt\Delta+1) \mathcal{E}_{M} Y_{(A}  \mathcal{P}_{B)}) \Pi_{\Delta-1} \nn
& \quad + ((\wt \Delta+1)  Y_A Y_M\mathcal{E}_{B} +\wt \Delta Y_B Y_M\mathcal{E}_{A}  )\Pi_{\Delta}  \nn
& \quad  +(\wt\Delta^2 Y_A Y_B \mathcal{E}_{M}-\wt\Delta \eta_{AB}\mathcal{E}_{M} - \eta_{AM}  \mathcal{E}_{B}) \Pi_{\Delta} \label{eq:covL1} \, .
\end {align}
Note that all of the operators in the identities above are normal ordered, so that $\cP$ always acts on scalar propagators before the spin-raising operator $\cE$. As before, the shadow relation $\cP^A\leftrightarrow \cX^A$, $\tDelta \leftrightarrow \Delta$ also applies in this case.

\paragraph{Spin two}

Similar identities hold for a spin-2 field. We have\footnote{This fixes some typos in Eq.~(C.1) of \cite{Li:2022tby}.}
\begin{align}
(\Pi_{\Delta})_{MN} &= \cE_M \cE_N \Pi_{\Delta} \,,\label{eq:spin2prop}\\
	\nabla_A (\Pi_{\Delta})_{MN} &= i\cE_M\cE_N \cP_A \Pi_{\Delta-1} -(\tDelta Y_A \cE_M\cE_N + Y_{(M}\cE_{N)}\cE_A) \Pi_\Delta\,,\\
	\nabla_A\nabla_B(\Pi_{\Delta})_{MN} &= -\cE_M\cE_N \cP_A \cP_B \Pi_{\Delta-2} - i\big(Y_{(M}\cE_{N)}\cE_{(A}\cP_{B)} +(\tDelta+1) \cE_M\cE_N Y_{(A}\cP_{B)}\big)\Pi_{\Delta-1}\nn
	&\quad +\big(2Y_{M}Y_{N} \cE_A\cE_B +(\tDelta+1) Y_AY_{(M} \cE_{N)}\cE_B+\tDelta Y_BY_{(M} \cE_{N)}\cE_A \big)\Pi_\Delta \nn
	&\quad +\big(\wt\Delta^2Y_AY_B\cE_M\cE_N- \tDelta\eta_{AB}\cE_M\cE_N-\eta_{B(M}\cE_{N)}\cE_A\big)\Pi_\Delta\,.\label{DDspin2}
\end{align}
Again, the equivalent expressions in $\cX$ representation can be easily obtained through the exchange $\cP^A\leftrightarrow \cX^A$, $\tDelta \leftrightarrow \Delta$. 
It is straightforward to generalize these identities to higher-spin cases.

\subsection{Bulk Derivation of $\LA JJ\O_\ell \RA$}
\label{sec:JJObulk}

As a demonstration, in this section we derive $\LA JJ\O_\ell\RA$ directly from the bulk. 
We focus on two simple cases with $\ell=0,2$, which are sufficient to illustrate the relevant qualitative features. 
Computing more complicated examples would be straightforward, but it would require the generalization of the propagator identities given in the previous section involving more covariant derivatives.

\subsubsection{$\LA JJ\O \RA$}
Consider first the case $\ell=0$. The cubic interaction between two photons $A^\mu$ and one scalar $\phi$ is
\begin{align}\label{FFphi}
S_{\rm int}  =  \int_{\text{AdS}_{d+1}} d^{d+1} x  \sqrt{g}\, \big(F_{\mu \nu} F^{\mu \nu} \phi \big)\,,
\end {align}
where $F_{\mu \nu} = \nabla_{[\mu} A_{\nu]}=\nabla_\mu A_\nu-\nabla_\nu A_\mu$  and we have suppressed a coupling constant.
For the time being, we will treat $A_\mu$ as a spin-1 field with a generic mass (massless or massive), but which has the same cubic interaction \eqref{FFphi}.
The three-point function for the above interaction is given by the following integral in embedding space:
\begin{align}
\langle J_{\Delta_J}J_{\Delta_J} \mathcal{O} \rangle &= \int dY\, \nabla_{[M} (\Pi_{1,\Delta_J})_{N]}\nabla^{[M}_{\phantom{\Delta_J}} \Pi_{2,\Delta_J}^{N]} \Pi_{3,\Delta}\,,
\label{eq:JJOEP}
\end{align}
where the covariant derivatives are with respect to the bulk coordinate $Y^M$. 
The subscript $J$ in $\Delta_J$ is a reminder that the spin-1 field has a generic scaling dimension $\Delta_J\ge d-1$ and is not necessarily conserved. 

We will rewrite the above integral as derivatives acting on external coordinates $X_i$ using identity \eqref{eq:covspin1},
in the form of weight-shifting operators acting on a scalar propagator. 
Making this substitution and using the following identities 
\beq
\begin{aligned}
    (Y\cdot \cE) \Pi_{\Delta}&=0\,,\\[-2pt] (Y\cdot \cP) \Pi_{\Delta-1} &= i\wt\Delta\Pi_{\Delta}\,,\\
    (Y\cdot\cE) \cP^M\Pi_{\Delta-1} &= i\cE^M \Pi_{\Delta}\,,
\end{aligned}\label{spin1id}
\eeq
we find that all $Y$-dependent pieces cancel among the terms in \eqref{eq:JJOEP}. We end up with 
\begin{equation}
\langle J_{\Delta_J} J_{\Delta_J} \mathcal{O} \rangle  \propto   (\EE{12}\PP{12}\, - :\!\DDt{12}\DDt{21}\!:) \LA \phi_{\Delta_J -1} \phi_{\Delta_J -1} \phi_{\Delta} \RA  - (\wt\Delta_J-1)^2  \EE{12} \langle \phi_{\Delta_J} \phi_{\Delta_J} \phi_{\Delta} \rangle \, ,
\end{equation}
where we have dropped a constant numerical prefactor, and the scalar three-point function is given by $\LA \phi_{\Delta_1} \phi_{\Delta_2} \phi_{\Delta_3} \RA = \int dY\, \Pi_{1,\Delta_1}\Pi_{2,\Delta_2}\Pi_{3,\Delta_3}$.  Notice that the second term above vanishes for conserved spin-1 gauge field with $\Delta_J = d-1$ or $\wt\Delta_J=1$, which then gives the $\ell=0$ version shown in~\eqref{eq:A1genlEP}. 

The basic procedure of bulk derivation in the $\cX$ representation is similar to that in the $\cP$ representation presented above, where all identities have 
$\cP\leftrightarrow\cX$, $\wt\Delta_J\leftrightarrow\Delta_J$, accompanied by a corresponding change of scaling dimensions in the scalar propagators. 
In the $\cX$ representation, we therefore end up with 
\begin{equation}
\langle J_{\Delta_J} J_{\Delta_J} \mathcal{O} \rangle  \propto  (\EE{12}\XX{12}\,-:\!\DD{12}\DD{21}\!:) \langle \phi_{\Delta_J+1} \phi_{\Delta_J+1} \phi_{\Delta} \rangle - (\Delta_J - 1)^2 \EE{12} \langle \phi_{\Delta_J} \phi_{\Delta_J} \phi_{\Delta} \rangle\,,
\end{equation}
giving the expression shown in \eqref{JJOPrep}.

\subsubsection{$\LA JJ\O_{\ell=2} \RA$}

There are two independent cubic interactions between two photons and one massive spin-$\ell$ particle with $\ell\ge 2$, shown in Table~\ref{tab:JJO}. 
We will denote the correlators derived directly from these two bulk Lagrangians as $\LA JJ\O_\ell\RA_{\cL_1}$ and $\LA JJ\O_\ell\RA_{\cL_2}$.

\paragraph{$\langle JJ \mathcal{O}_2 \rangle_{\cL_1}$} We first consider the cubic action
\begin{align}
S_{\text{int}} 
=  \int_{\text{AdS}_{d+1}} d^{d+1} x  \sqrt{g} \big(\nabla_\alpha F_{\mu \nu} \nabla_\beta F^{\mu \nu} S^{\alpha\beta} \big)\,.
\end{align}
The corresponding three-point function $\LA J J  \O_2 \RA_{\cL_1}$ is computed from the integral
\begin{align}
\langle JJ \mathcal{O}_2 \rangle_{\cL_1} = \int d Y\, \nabla_A \nabla_{[M} (\Pi_{1, \Delta_J})_{N]}\nabla_B \nabla^{[M}\Pi_{2, \Delta_J}^{N]} \Pi_{3,\Delta}^{AB}\,,
\label{eq:JJO2l1}
\end{align}
where we have again assumed a generic scaling dimension $\Delta_J\ge d-1$ for the spin-1 field.
The basic procedure is similar to that of the $\ell=0$ case, except that one needs a different set of relations to simplify the intermediate steps in the bulk integral.  
We can replace the double covariant derivatives acting on the spin-1 propagator using the identities \eqref{eq:covL1}, \eqref{spin1id}, as well as
\beq
\begin{aligned}
 Y^M  (\Pi_{\Delta})_{MN} &= 0 \,,\\
 (Y\cdot \cE) \cP^M \cP^N \Pi_{\Delta-2} &= i \cE^{(M}\cP^{N)} \Pi_{\Delta-1} \,.
\end{aligned}
\eeq
After many cancellations of the $Y$-dependent terms, we arrive at the following differential representation of the correlator:
\begin{align}
\LA \wh{J_{\Delta_J}J_{\Delta_J}\O_2}\RA_{\cL_1} &\propto ( \EE{12}\PP{12} \, -  :\!\DDt{12}\DDt{21}\!: )\DDt{31}\DDt{32} -\Big[\EE{13}\EE{23}\PP{12} + \EE{12}\DDt{31}\DDt{32}((\tDelta_J+1)^2-3 )  \nn
&\quad - (2\tDelta_J-1) (\EE{13}\DDt{21}\DDt{32} + 1\leftrightarrow 2)\Big] - (\tDelta_J-1)^2\EE{13}\EE{23} \, .
\end{align}
For the conserved case $\Delta_J=d-1$, the last term disappears, and the result is given by a linear combination of the expressions obtained via directly uplifting amplitudes---\eqref{eq:A1genlEP} and \eqref{eq:A2genlEP}---as follows:
\begin{align}
    \LA JJ\O_2\RA_{\cL_1} \propto \LA JJ\O_2\RA^{(1)}+ \LA JJ\O_2\RA^{(2)}\,.
\end{align}
This means that the directly-uplifted representation \eqref{JJO2gen1} in fact mixes terms with different derivative orders.

\paragraph{$\langle JJ \mathcal{O}_2 \rangle_{\cL_2}$} 

The other independent cubic action is
\begin{align}
S_{\text{int}} & = \int_{AdS_{d+1}} d^{d+1} x \sqrt{g} \big(F_{\mu \alpha} F^{\mu\beta} {S^{\alpha}}_\beta\big)\,. 
\end {align}
The integral for the corresponding three-point function is
\begin{align}
\langle JJ \mathcal{O}_2 \rangle_{\cL_2} = \int dY\, \nabla_{[M} (\Pi_{1, \Delta_J})_{A]} \nabla^{[M}(\Pi_{2,\Delta_J})^{B]} {(\Pi_{3,\Delta})^{A}}_B\,.
\label{eq:JJO2l2}
\end{align}
As before, we assume $\Delta_J\ge d-1$ in the derivation. Using the propagator identities, we obtain the differential representations
\begin{align}
&\LA \wh{J_{\Delta_J} J_{\Delta_J} \mathcal{O}_2} \RA_{\cL_2} \propto  \EE{13}\EE{23} \cP_{12} + \EE{12} \DDt{31} \DDt{32} - \EE{23}\DDt{12}\DDt{31} - \EE{13} \DDt{21}\DDt{32}   - (\tDelta_J-1)^2 \EE{13} \EE{23}  \nn[2pt]
&\qquad\qquad =   \EE{13} \EE{23} \cX_{12} + \EE{12} \DD{31} \DD{32} - \EE{23}\DD{12}\DD{31} - \EE{13} \DD{21} \DD{32} - (\Delta_J - 1)^2 \EE{13}\EE{23}\, ,
\end{align}
where we have used the shadow relation in the second line. 
For the conserved case $\Delta_J = d-1$ or $\tDelta=1$, the last term in the $\cP$ representation above vanishes, reproducing \eqref{eq:A2genlEP}, whereas the second line gives the $\cX$ representation shown in \eqref{eq:A2genldown}.

\section{Comparison with Helicity Basis}\label{app:pairing}

It is instructive to compare the amplitude basis discussed in this paper with the helicity basis constructed in \cite{Caron-Huot:2021kjy, Li:2021snj}. 
In the latter basis, the OPE data for conserved currents in mean field theory were found to be diagonal in $d=3$, which suggests that the structures are well chosen. 
We first present the generalization of the helicity basis in $d \geq 3$, and then make a comparison with the amplitude basis.

The core idea is to construct SO($d-1$) symmetric tensor structures perpendicular to momentum $k$ conjugate to $x$ in the conformal frame $(x_1,x_2,x_3)=(0, x, \infty)$. 
One then organizes the structures through the little group SO($d-1$) that leaves $k$ invariant. 
The construction of the polarization tensor structures that are perpendicular to $k$ can be achieved by the following projection operator\footnote{This fixes some typos in Eq.~(5.3) of~\cite{Li:2021snj}.}
\begin{align}
	\cP_\epsilon^{(\ell)}&=\cP_\epsilon^{(\ell,\ell)}\cP_\epsilon^{(\ell-1,\ell)} \cdots  \cP_\epsilon^{(0,\ell)}\,,\\
	\cP_\epsilon^{(m,\ell)}&\equiv 1-\frac{2}{(d-4+\ell+m)(\ell-m+1)}\frac{\epsilon\cdot k}{k^2}(k \cdot T_\epsilon)\,,\quad \cP_\epsilon^{(0,\ell)}\equiv1\,,
\end{align}
where $T_\epsilon$ is the Todorov operator defined in~\eqref{eq:Todorov}. This operator is constructed such that it is transverse to $k$:
\begin{align}
	(k \cdot \cD_\epsilon) \cP_\epsilon^{(\ell)} \epsilon^{\mu_1}\cdots \epsilon^{\mu_\ell}=0\,.
\end{align}
The parity-even tensor structures can then be constructed as~\cite{Li:2021snj}
\begin{align}
    \mathbb{H}_{\O_1\O_2\O_3}^{i_1i_2i_3} &\propto (\epsilon_1\cdot k)^{\ell_1-i_1}(\epsilon_2\cdot k)^{\ell_2-i_2}(\epsilon_3\cdot k)^{\ell_3-i_3}k^{\alpha-d}\nn
	&\quad \times\cP_{\epsilon_1}^{(i_1)}\cP_{\epsilon_2}^{(i_2)}\cP_{\epsilon_2}^{(i_3)}(\epsilon_1\cdot\epsilon_2)^{\frac{i_{123}}{2}}(\epsilon_2\cdot\epsilon_3)^{\frac{i_{231}}{2}}(\epsilon_3\cdot\epsilon_1)^{\frac{i_{312}}{2}}\,,
\end{align}
where $\alpha\equiv (\Delta_{1}+\Delta_2-\Delta_3)-(\ell_1-i_1)-(\ell_2-i_2)-(\ell_3-i_3)$ and $i_{abc} = i_a + i_b - i_c$. 
For $i_1,i_2$ that run from 0 to $\ell_1,\ell_2$, respectively, the third index $i_3$ takes values $|i_1-i_2|,|i_1-i_2|+2,\cdots,i_1+i_2$. 
For $\ell_1 = \ell_2 = \ell$, there are a total of $(\ell+1)$ tensor structures, which form a basis for parity-even three-point structures with two conserved spin-$\ell$ tensors. 
We will drop the overall normalization factors, which are unimportant for our purposes.

By construction, the helicity basis structures are orthogonal under the three-point inner product~\cite{Karateev:2018oml}
\begin{align}
\Big(\langle \mathcal{O}_1 \mathcal{O}_2 \mathcal{O}_3 \rangle^{(a)}, \langle \wt{\mathcal{O}}_1 \wt{\mathcal{O}}_2 \wt{\mathcal{O}}_3 \rangle^{(b)} \Big)& = \int \frac{d^d x_1d^d x_2 d^d x_3} {\text{vol}\, \text{SO}(d+1, 1)} \langle \mathcal{O}_1 \mathcal{O}_2 \mathcal{O}_3 \rangle^{(a)} \langle \wt{\mathcal{O}}_1  \wt{\mathcal{O}}_2 \wt{\mathcal{O}}_3 \rangle^{(b)}\phantom{\,.}\nn
&=\frac{\langle \mathcal{O}_1 (0) \mathcal{O}_2 (e) \mathcal{O}_3 (\infty) \rangle^{(a)}  \langle \wt{\mathcal{O}}_1 (0) \wt{\mathcal{O}}_2 (e) \wt{\mathcal{O}}_3 (\infty) \rangle^{(b)}}{2^d\, \text{vol}\, \text{SO}(d+1, 1)}\,,
\end{align}
where we have suppressed the contraction of tensor indices. In the second line we have gauge fixed the conformal symmetry to the conformal frame, with $2^{-d}$ being the Faddeev-Popov determinant, and $e$ is a unit vector. 
The operator insertion at $\infty$ is defined as the limit $\O(\infty)=\lim_{r\to\infty}r^{2\Delta}\O(r)$, which gives a finite correlation function.

\paragraph{Spin one} 

For conserved spin-1 currents, the parity-even helicity basis in momentum space takes the form~\cite{Caron-Huot:2021kjy}
\begin{align}\label{HJJO}
    \begin{bmatrix}\mathbb{H}_{JJ\mathcal{O}_\ell}^{(1)}\\[2pt] \mathbb{H}_{JJ\mathcal{O}_\ell}^{(2)} \end{bmatrix} = \begin{bmatrix}
t_{12} \\[2pt]     \frac{t_{13}t_{23}}{(\eps_3\cdot k)^2}-\frac{t_{12}}{1-d}
    \end{bmatrix} \frac{(\eps_3\cdot \hat k)^{\ell_3}}{k^{\Delta_3-d+4}}\,,
\end {align}
where $t_{ab}\equiv k^2(\epsilon_a\cdot\epsilon_b)-(\epsilon_a\cdot k)(\epsilon_b\cdot k)$ and we have labeled the structures as $\{\mathbb{H}_{JJ\O_\ell}^{(1)},\mathbb{H}_{JJ\O_\ell}^{(2)}\}=\{\mathbb{H}_{JJ\O_\ell}^{110},\mathbb{H}_{JJ\O_\ell}^{112}\}$. Both of these structures are transverse with respect to $\epsilon_1$ and $\epsilon_2$, and correspond to SO($d-1$) traceless symmetric tensors with respect to $\epsilon_3$. 
The three-point pairing matrix between these two structures is diagonal:
\begin{align}
\Big(\mathbb{H}_{ JJ \O_\ell}^{(a)},\, \wt{\mathbb{H}}_{JJ \O_\ell}^{(b)} \Big)\propto \begin{bmatrix}
1 & 0 \\[2pt]
0 &  \frac{ (d-2)(d-2+\ell) (d-1+\ell)}{ (d-1)^2 \ell (\ell-1)} \\
\end{bmatrix}_{ab}\, ,
\end {align}
where $\wt{\mathbb{H}}_{\O_1\O_2\O_3}^{(a)}\equiv \mathbb{H}_{\wt\O_1\wt\O_2\wt\O_3}^{(a)}$ denotes the shadow structures. 
The lower diagonal entry is singular for $\ell=0,1$, which reflects the fact that the second structure only exists for $\ell_3 \ge 2$ for generic $\Delta_3$.

After constructing the structures that are orthogonal to $k$, it is convenient to Fourier transform these back to position space. 
There are four independent tensor structures in the conformal frame symmetric in $z_1$ and $z_2$, given by
\begin{align}
q_1 &=  \red{z_{12}}\,, \hskip -30pt&& \hskip -30pt q_3 = \green{(z_{13}v_2+z_{23}v_1)v_3^{-1} }\,,\nn
q_2 &=  \red{v_1v_2} \,, \hskip -30pt&&\hskip -30pt q_4 = \blue{z_{13}z_{23}v_3^{-2}}\, ,
\end{align}
where we have defined $z_{ij} \equiv z_i \cdot z_j$ and  $v_i \equiv z_i \cdot \hat x$.  
We can then express the helicity basis structures \eqref{HJJO} in position space as
\begin{align}
\begin{bmatrix}
\mathbb{H}_{JJ\mathcal{O}_\ell}^{(1)} \\[2pt]
\mathbb{H}_{JJ\mathcal{O}_\ell}^{(2)}
\end{bmatrix} = M^{JJ\mathcal{O}_\ell} \begin{bmatrix}
	q_1\\[-2pt] \vdots\\[-2pt] q_4
\end{bmatrix} \frac{v_3^{\ell_3}} {|x|^{\Delta_1 + \Delta_2 - \Delta_3 }}\, ,\label{matrixMB}
\end {align}
where $M_{JJ\mathcal{O}_\ell}$ is the $2 \times 4$ matrix given by~\cite{Li:2021snj} 
\begin{align}
M_{JJ\O_\ell} \propto \begin{bmatrix}
2n(d-\beta-1) & 4n(n-1) & 2n\ell & \ell(\ell-1)\\
-\frac{2n\beta}{d-1} & -\frac{4(d-2)n(n-1)}{d-1} & 2n(\frac{\ell + (d-1)(2-\wt\Delta)}{d-1}) & \frac{\cJ-(d-1)(\wt\Delta-2)^2}{d-1}
\end{bmatrix},
\end{align}
with $\cJ=\ell(\ell+d-2)$, $\Delta-\ell=2(d-2+n)$ and $\beta=\Delta+\ell$. 
In particular, when $\Delta_3 = d-1$ and $\ell_3 = 1$, the last column of $M_{JJ\O_\ell}$ vanishes. 
In this case, the helicity basis is related to the two three-point functions $\LA JJJ\RA_{F^3}$ and $\LA JJJ\RA_{\rm YM}$, shown in \eqref{eq:JJJYMup} and \eqref{eq:JJJF3up}, as
\begin{align}
    \begin{bmatrix}
      \LA JJJ\RA_{F^3}\\[4pt] \LA JJJ\RA_{\rm YM}
    \end{bmatrix} \propto \begin{bmatrix}
        d(d-2) & 0  \\[4pt] 3-d & 1-d
    \end{bmatrix} \begin{bmatrix}
       \mathbb{H}_{JJJ}^{(1)}\\[2pt] \mathbb{H}_{JJJ}^{(2)}
    \end{bmatrix} .
\end{align}
When $d=3$, we see that the helicity basis precisely corresponds to the amplitude structures, as was shown in \cite{Caron-Huot:2021kjy}.

For the amplitude basis $\langle JJ\mathcal{O}_\ell \rangle^{(a)}$ with non-conserved operators, given in \eqref{eq:A1genldown} and \eqref{eq:A2genldown}, the two bases are related by the following change-of-basis matrix:
\begin{align}
    \begin{bmatrix}
         \LA JJ\O_\ell\RA^{(1)}\\[3pt]\LA JJ\O_\ell\RA^{(2)}
    \end{bmatrix} \propto \begin{bmatrix}
          \frac{- \Delta(\Delta+d-4)}{2}+\frac{\ell (d-3)(d+\ell-2)} {2(d-1)}  & \ell (\ell-1) \\[2pt]  - \frac{(d-2)(d-3)(d+\Delta+\ell-2)}{(d-1)(\Delta+\ell-d+2)} & - \frac{2 (d-2) (\ell + \Delta-1)}{(\ell + \Delta - d+2)} 
    \end{bmatrix} \begin{bmatrix}
         \mathbb{H}_{JJ\O_\ell}^{(1)}\\[3pt] \mathbb{H}_{JJ\O_\ell}^{(2)}
    \end{bmatrix} .
\end{align}
The unitarity bound $\Delta \geq d - 2 + \ell$ guarantees that the coefficients have positive denominators, and the matrix is  invertible for $d \geq 3$ and $\ell \geq 2$. In $d=3$, the second structure $\LA JJ\O_\ell\RA^{(2)}$ becomes proportional to $\mathbb{H}_{JJ\O_\ell}^{(2)}$.

\paragraph{Spin two} 

For stress tensors, the parity-even helicity basis takes the form
\begin{equation}
   \hskip -10pt \begin{bmatrix}\label{HTTO}
        \mathbb{H}_{TT\mathcal{O}_\ell}^{(1)}\\[2pt]
        \mathbb{H}_{TT\mathcal{O}_\ell}^{(2)}\\[2pt]
        \mathbb{H}_{TT\mathcal{O}_\ell}^{(3)}
    \end{bmatrix}
    = \begin{bmatrix}
     t_{12}^2-\frac{(\epsilon_1 \cdot k)^2(\epsilon_2 \cdot k)^2}{d-1}\\
 \frac{(d-1)t_{12}t_{23}t_{13} + (\epsilon_3 \cdot k)^2t_{12}^2 + (\epsilon_1 \cdot k)^2 t_{23}^2 + (\epsilon_2 \cdot k)^2 t_{13}^2}{(d-1)(\epsilon_3 \cdot k)^2} - \frac{2 (\epsilon_1 \cdot k)^2 (\epsilon_2 \cdot k)^2}{(d-1)^2}\\
	\frac{t_{13}^2t_{23}^2}{(\eps_3\cdot k)^4}{+}\frac{4t_{12}t_{23}t_{13} + \frac{2}{d+1} (\epsilon_3 \cdot k)^2t_{12}^2 - (\epsilon_1 \cdot k)^2 t_{23}^2 - (\epsilon_2 \cdot k)^2 t_{13}^2}{(d+3)(\epsilon_3 \cdot k)^2}{+}\frac{(\epsilon_1 \cdot k)^2(\epsilon_2 \cdot k)^2}{(d+3)(d+1)}
    \end{bmatrix} 
    \frac{(\eps_3\cdot \hat k)^{\ell_3}}{k^{\Delta_3-d+4}}\,,
\end{equation}
where we have labeled $\{\mathbb{H}_{TT\mathcal{O}_\ell}^{(1)},\mathbb{H}_{TT\mathcal{O}_\ell}^{(2)},\mathbb{H}_{TT\mathcal{O}_\ell}^{(3)}\} = \{\mathbb{H}_{TT\mathcal{O}_\ell}^{220},\mathbb{H}_{TT\mathcal{O}_\ell}^{222},\mathbb{H}_{TT\mathcal{O}_\ell}^{224}\}$. 
Again, all of these structures are transverse with respect to $\epsilon_1$ and $\epsilon_2$, and correspond to SO($d-1$) traceless symmetric tensors with respect to $\epsilon_3$. 
As in the spin-1 case, the three-point pairing of the helicity basis structures is diagonal by construction:
\begin{align}
\Big(\mathbb{H}_{ TT \O_\ell}^{(a)},\, \wt{\mathbb{H}}_{ TT \O_\ell}^{(b)} \Big)\propto \begin{bmatrix}
1 & 0 & 0 \\
0 & \frac{(d^2 - 9) (d+\ell - 2) (d+\ell-1)} {2 \ell (\ell-1) (d-1)^2 (d+1)} & 0 \\
0 & 0 & \frac{2 d (d+\ell+1) (d+\ell)(d+\ell-1) (d+\ell-2)}{ \ell (\ell-1)(\ell-2) (\ell-3) (d+3)(d+1)^2} \\
\end{bmatrix}_{ab} \, .
\end{align}
From these entries, we see that the structures $\mathbb{H}_{TT\cO_\ell}^{(2)}$ and $\mathbb{H}_{TT\cO_\ell}^{(3)}$ only exist for $\ell \ge 2$ and $\ell \geq 4$, respectively.

Back in position space within the conformal frame, the 10 independent tensor structures that are symmetric in 1 and 2 are given by
\beq
\begin{aligned}
q_1 & = \red{z_{12}^2}\,, \quad && q_6 = \green{z_{12} z_{13} z_{23} v_3^{-2}} \,,\\
q_2 & = \red{z_{12} v_1 v_2}\,, \quad && q_7 = \green{z_{13} z_{23} v_1 v_2 v_3^{-2}} \,,\\
q_3 & = \red{v_1^2 v_2^2}\,, \quad && q_8 = \green{(z_{23}^2 v_1^2 + z_{13}^2 v_2^2) v_3^{-2}}\,, \\
q_4 & = \orange{z_{12} (z_{23} v_1 + z_{13} v_2) v_3^{-1}}\,, \quad && q_9 = \blue{(z_{13} v_2 + z_{23} v_1 ) z_{13} z_{23} v_3^{-3} }\,,\\
q_5 & = \orange{(z_{13} v_2 + z_{23} v_1)v_1v_2 v_3^{-1}}\,, \quad && q_{10} = \purple{z_{13}^2 z_{23}^2 v_3^{-4} }\,.
\end{aligned}
\eeq
The spin-2 helicity basis \eqref{HTTO} can then be expressed as
\begin {align}
\begin{bmatrix}
\mathbb{H}_{TT\mathcal{O}_\ell}^{(1)} \\[2pt]
\mathbb{H}_{TT\mathcal{O}_\ell}^{(2)} \\[2pt]
\mathbb{H}_{TT\mathcal{O}_\ell}^{(3)}
\end {bmatrix} = M_{TT\mathcal{O}_\ell} \begin{bmatrix}
	q_1\\[-2pt] \vdots\\[-2pt] q_{10}
\end{bmatrix} \frac{v_3^{\ell_3}} {|x|^{\Delta_1 + \Delta_2 - \Delta_3 }} \,,
\end {align}
where $M_{TT\mathcal{O}_{\ell}}$ is a $3 \times 10$ matrix. 
The helicity basis is related to the amplitude basis $\LA TT\cO_\ell \RA^{(a)}$ constructed in the main text by a change of basis as
\begin{align}
    \begin{bmatrix}
        \LA TT\mathcal{O}_\ell\RA^{(1)}\\[2pt]
        \LA TT\mathcal{O}_\ell\RA^{(2)}\\[2pt]
        \LA TT\mathcal{O}_\ell\RA^{(3)}
    \end{bmatrix} = R_{TT\O_\ell}  \begin{bmatrix}
        \mathbb{H}_{TT\mathcal{O}_\ell}^{(1)}\\[2pt]
        \mathbb{H}_{TT\mathcal{O}_\ell}^{(2)}\\[2pt]
        \mathbb{H}_{TT\mathcal{O}_\ell}^{(3)}
    \end{bmatrix} .
\end{align}
Due to their large sizes, we show the explicit expressions for the matrices $M_{TT\O_\ell}$ and $R_{TT\O_\ell}$ in the \textsc{Mathematica} notebook~\href{https://github.com/haydenhylee/ampbasis}{\faGithub}.

\newpage
\bibliographystyle{JHEP}
\bibliography{Refs}

\end{document}